\documentclass[12pt,a4paper]{article}
\usepackage[utf8]{inputenc}
\usepackage{epsfig}
\usepackage{mathtools}
\usepackage{amsfonts}
\usepackage{amssymb}
\usepackage{amsmath}
\usepackage{cancel}
\usepackage{color}
\usepackage{slashed}
\usepackage{cite}
\usepackage{caption}
\usepackage{subcaption}
\usepackage{comment}
\usepackage{marginnote}

\usepackage{tikz}
\usetikzlibrary{shapes}
\usetikzlibrary{trees}
\usetikzlibrary{calc}
\usetikzlibrary{decorations.pathmorphing}
\usetikzlibrary{decorations.markings}

\def\bP{{\bf P}}
\def\bnu{{\bar{\nu}}}
\def\sign{\text{sgn}}

\def\s{\sigma}

\def\bp{{\bf p}}
\def\btp{{\bf\tilde p}}
\def\btP{{\bf\tilde P}}
\def\bP{{\bf P}}
\def\OO{\mathcal{O}}
\def\Z{\mathcal{Z}}
\def\P{\mathcal{P}}

\newcommand{\barH}{{\bar H}}
\newcommand{\egamma}{\gamma_E}

\begin{document}

\begin{center}

\vspace{1cm}

{ \Large\bf ABJM quantum spectral curve and Mellin transform} \vspace{1cm}

{\large R.N. Lee$^{1}$ and  A.I. Onishchenko$^{2,3,4}$}\vspace{0.5cm}

{\it $^1$ Budker Institute of Nuclear Physics, Novosibirsk,
	Russia,\\
	$^2$Bogoliubov Laboratory of Theoretical Physics, Joint
	Institute for Nuclear Research, Dubna, Russia, \\
	$^3$Moscow Institute of Physics and Technology (State University), Dolgoprudny, Russia\\
	$^4$Skobeltsyn Institute of Nuclear Physics, Moscow State University, Moscow, Russia}\vspace{1cm}

\abstract{The present techniques for the perturbative solution of quantum spectral curve problems in $\mathcal{N}=4$ SYM and ABJM models
are limited to the situation when the states quantum numbers are given explicitly as some integer numbers.
These techniques are sufficient to recover full analytical structure of the conserved charges provided that we know a finite basis of functions in terms of which they could be written explicitly. It is known that in the case of $\mathcal{N}=4$ SYM both the contributions of asymptotic Bethe ansatz and wrapping or finite size corrections are expressed in terms of the harmonic sums. However, in the case of ABJM model only the asymptotic contribution can still be written in the harmonic sums basis, while the wrapping corrections part can not. Moreover, the generalization of harmonic sums basis for this problem is not known. In this paper we present a Mellin space technique for the solution of multiloop Baxter equations, which is the main ingredient for the solution of corresponding  quantum spectral problems, and provide explicit results for the solution of ABJM quantum spectral curve in the case of twist 1 operators in $sl(2)$ sector for arbitrary spin values up to four loop order with explicit account for wrapping corrections. It is shown that the result for anomalous dimensions could be expressed in terms of harmonic sums decorated by the fourth root of unity factors, so that maximum transcendentality principle holds.}
\end{center}

\begin{center}
	ABJM, Baxter equations, Mellin transform	
\end{center}

\newpage

\tableofcontents{}\vspace{0.5cm}

\renewcommand{\theequation}{\thesection.\arabic{equation}}

\section{Introduction}

In the past time following the discovery of AdS/CFT duality \cite{tHooftDuality,MaldacenaAdSCFT,GKP,WittenAdSHolography}
we faced a lot of progress in the study of integrable structures behind the quantum field theories with extended supersymmetry in dimensions greater then two, see for a review and introduction \cite{IntegrabilityReview,IntegrabilityPrimer,IntegrabilityDeformations,IntegrabilityDefects,IntroductionQSC,IntegrabilityStructureConstants}. The most well understood theories are given by $\mathcal{N}=4$ SYM in four and $\mathcal{N}=6$ super Chern-Simons theory in three dimensions. The latter theory is more known as ABJM model \cite{ABJM1}. In particular, different techniques from the world of integrable systems, such as worlsheet and spin-chain S-matrices \cite{StaudacherSMatrix,BetheAnsatzQuantumStrings,BeisertDynamicSmatrix,BeisertAnalyticBetheAnsatz,TranscedentalityCrossing,JanikWorldsheetSmatrix,ArutyunovFrolovSmatrix,ZamolodchikovFaddevAlgebraAdS,N6Smatrix}, Asymptotic Bethe Ansatz (ABA) \cite{MinahanZarembo,N4SuperSpinChain,LongRangeBetheAnsatz,TranscedentalityCrossing,MinahanZaremboChernSimons,SpinChainsN6ChernSimons,AllloopAdS4} and Thermodynamic Bethe Ansatz (TBA) \cite{TBAN4,TBAN4proposal,TBAexcitedstates,TBAMirrorModel} as well as $Y$ and $T$-systems \cite{YsystemAdS5,TBAfromYsystem,WronskianSolution,SolvingYsystem,TBAYsystemAdS4,GromovYsystemAdS4,DiscontinuityRelationsAdS4} were shown to be very useful for the computation of conformal spectrum of these theories. The integrability based methods were also applied to the study of quark-antiquark potential \cite{potentialTBA,IntegrableWilsonLoops,cuspQSC,potentialQSC}, expectation values of polygonal Wilson loops at strong coupling and beyond \cite{BubbleAnsatz,YsystemScatteringAmplitudes,OPEpolygonalWilsonLines,SmatrixFiniteCoupling,OPEHelicityAmplitudes,AsymptoticBetheAnsatzGKPvacuum}, eigenvalues of BFKL kernel \cite{adjointBFKL,GromovBFKL1,GromovBFKL2}, structure constants \cite{StructureConstantsPentagons,StructureConstantsWrappingOrder,ClusteringThreePointFunctions,StructureConstantsLightRayOperators} and one-point functions of operators in the defect conformal field theory \cite{defectCFT1,defectCFT2,defectCFT3}.

Recently, the detailed study of TBA equations for  $\mathcal{N}=4$ SYM and ABJM models resulted in the discovery of very effective Quantum Spectral Curve (QSC) formulation for these models \cite{N4SYMQSC1,N4SYMQSC2,twistedN4SYMQSC,N4SYMQSC3,ABJMQSC,ABJMQSCdetailed,QSCetadeformed}. The latter gives an alternative reformulation of TBA equations as a nonlinear Riemann-Hilbert problem. The iterative procedure for perturbative solution of the mentioned Riemann-Hilbert problems for these theories at weak coupling  was finally proposed in \cite{VolinPerturbativeSolution,ABJMQSC12loops}. However, the presented technique is limited to the situation when the states quantum numbers are given explicitly as some integer numbers. It is sufficient for the recovery of full analytical structure of the conserved charges in the respective spin chains provided that we know a finite basis of functions in terms of which they could be written explicitly. It is known that in the case of $\mathcal{N}=4$ SYM both the contributions of asymptotic Bethe ansatz and wrapping or finite size corrections for twist-2 and twist-3 operators are expressed in terms of the harmonic sums \cite{N4SYM1loop,N4SYM2loop1,N4SYM2loop2,N4SYM2loop3,N4SYM2loop4,N4SYM3loop,N4SYM4loop,N4SYM5loop,N4SYM6looptwist3,N4SYM6loop,N4SYM7loop}. However, in the case of ABJM model it was possible to express in the harmonic sums basis only the asymptotic contribution, while the wrapping corrections part misses such a representation. Moreover, the generalization of harmonic sums basis for this problem is not known at present \cite{Beccaria1,MinahanABJM4loop1,MinahanABJM4loop2,ABJMspectroscopy,ABJMsuperspace,Beccaria2,ABJMQSC12loops}. In this paper we present a Mellin space technique for the solution of multiloop Baxter equations, which is the main ingredient for the solution of corresponding  quantum spectral problems, and provide explicit results for the solution of ABJM quantum spectral curve in the case of twist 1 operators in $sl(2)$ sector for arbitrary spin values up to four loop order with explicit account for wrapping corrections. The results for anomalous dimensions we obtained could be further expressed in terms of harmonic sums decorated by the fourth root of unity factors. 

The present paper is organized as follows.  In the next section we remind the reader the formulation of ABJM quantum spectral curve. Section \ref{PerturbativeSolution} contains the solution of ABJM QSC in the case of twist 1 operators in $sl(2)$ sector up to four loop order. Next, in section \ref{BaxterEquationsMellin} we present the details of solution of corresponding Baxter equations with Mellin space technique. Section \ref{AnomalousDimensions} contains the discussion of our results for anomalous dimensions of twist 1 operators up to four loop order  and for arbitrary values of spin variable. Finally, in section \ref{Conclusion} we come with our conclusion. Appendices and {\it Mathematica} notebooks contain some details of our calculation.

\section{ABJM quantum spectral curve}\label{ABJM-QSC}

ABJM is a three-dimensional $\mathcal{N}=6$ Chern-Simons theory with product gauge group $U(N)\times \hat{U}(N)$ at levels $\pm k$. The field content of the theory is given by two gauge fields $A_{\mu}$ and $\hat{A}_{\mu}$, four complex scalars $Y^A$ and four Weyl spinors $\psi_A$. The matter fields transform in the bi-fundamental representation of the gauge group. The global symmetry group of ABJM theory for Chern-Simons level $k > 2$ is given by orthosymplectic supergroup $\text{OSp} (6|4)$ \cite{ABJM1,ABJM2} and the ``baryonic'' $U(1)_b$ \cite{ABJM2}. In the present paper we will be interested in anomalous dimensions of $sl (2)$-like states given by single-trace operators of the form \cite{KloseABJMreview}:
\begin{eqnarray}
\text{tr} \left[ D_{+}^S (Y^1 Y_4^\dagger)^L\right] . \label{sl2op}
\end{eqnarray}
In fact in what follows we will restrict ourselves with only twist 1 ($L=1$) operators. It should be noted that these operators characterized by Dynkin labels $[L+S,S;L,0,L]$ do not form a closed subsector by themselves but belong to a wider $\text{OSp}(2|2)$ subsector. At the moment the most advanced method to deal with spin chain spectral problems arising in the study if $\text{AdS}_{d+1}/\text{CFT}_d$ duality is offered by quantum spectral curve (QSC) method. For the case of ABJM theory QSC formulation was introduced in \cite{ABJMQSC,ABJMQSCdetailed}, see also \cite{ABJMQSC12loops}. 

The QSC method is another reformulation of Thermodynamic Bethe Ansatz (TBA) as a set of functional equations, such as $Y$ or $T$-systems. The QSC method or ${\bf P\mu}$-system is special in a sense, that it involves only a finite number of objects satisfying a set of nonlinear matrix Riemann-Hilbert equations.  The ${\bf P\mu}$-system for ABJM consists of six functions ${\bf P}_a, a=1,\ldots ,6$ and an antisymmetric $6\times 6$ matrix $\mu_{a b}$. Both ${\bf P}_a$ and $\mu_{a b }$ are functions of spectral parameter $u$. The ${\bf P}_a$ functions are defined on a Riemann sheet with a single cut running from  $-2 h$ to $+2 h$ ($h$ is ABJM QSC coupling constant), while functions $\mu_{a b}$ have an infinity of branch cuts at intervals $(-2h,+2h) + i n, n\in \mathbb{Z}$ and satisfy a simple relation
\begin{eqnarray}
\tilde{\mu}_{a b} (u) = \mu_{a b} (u+i) ,
\end{eqnarray}
where $\tilde{f}$ here and in the following will denote a function $f$ analytically continued around one of the branch points on the real axis. It is important to mention that, in contrast to $\mathcal{N}=4$ SYM, ABJM QSC coupling constant $h$ is a nontrivial function of ABJM t'Hooft coupling constant $\lambda$ \cite{SpinChainsN6ChernSimons,StringDualN6ChernSimons}, which scales as $h\sim\lambda$ at small and as $h\sim\sqrt{\lambda /2}$ at strong coupling constant. An important conjecture for the exact form of $h(\lambda)$ was made in \cite{hconjecture1,hconjecture2} by a comparison with the structure of localization results. The functions ${\bf P}_a$ and $\mu_{a b }$ satisfy the set of nonlinear constraints
\begin{eqnarray}
\bP_5 \bP_6 &=& 1 + \bP_2 \bP_3 - \bP_1 \bP_4, \label{Pconstr}\\
\mu \chi \mu \chi &=& 0 , \label{muconstr}
\end{eqnarray}
where nonzero entries of $6\times 6$ symmetric matrix $\chi$ are given by
\begin{eqnarray}
\chi^{14} = \chi^{41} = -1 , \;\;\; \chi^{23} = \chi^{32} =  1 , \;\;\; \chi^{56} = \chi^{65} = -1 .
\end{eqnarray}
The fundamental Riemann-Hilbert relations for ${\bf P}_a$ and $\mu_{a b }$ functions are written as 
\begin{eqnarray}
\tilde { \bP }_a &=& { \bP }_a - \mu_{ab}\chi^{bc} { \bf P }_c , \label{Pmueqn1}\\
\mu_{ab}-\tilde \mu_{ab}&=&- { \bf P }_a\tilde { \bf P }_b + { \bf P }_b\tilde { \bf P }_a . \label{Pmueqn2}
\end{eqnarray}
It should be noted that, similar to the $\mathcal{N}=4$ SYM there is a complementary set of functions satisfying their Riemann-Hilbert relations, so called ${\bf Q}\omega$-system \cite{ABJMQSC}. The ${\bf Q}\omega$-system is similar to ${\bf P}\mu$-system (\ref{Pconstr})-(\ref{Pmueqn2}) with replacements
\begin{eqnarray}
\bP_a \rightarrow {\bf Q}_a\, , \hspace{1cm} \mu_{ab} \rightarrow\omega_{ab} \,.
\end{eqnarray}
The ${\bf Q}$ and $\omega$ functions have different cut structure however, see for details \cite{ABJMQSC,ABJMQSCdetailed} and \cite{N4SYMQSC1,N4SYMQSC2}. In the case of ABJM model it is convenient to parametrize $\mu_{a b}$ matrix in terms of 8 functions 
$\nu_i, \bar{\nu}_i, i = 1,\dots ,4$ as \cite{ABJMQSC,ABJMQSCdetailed}:
\begin{equation}
\mu_{a b}=
\left(
\begin{array}{cccccc}
0 & \nu_1 \bnu_{1} & \nu_2 \bnu_{2} &  \bnu_2 \nu_{3}-\bnu_1 \nu_{4} & \nu_1 \bnu_2 & \bnu_1 \nu_2 \\
-\nu_1 \bnu_{1} & 0 & \bnu_2 \nu_3 + \nu_1 \bnu_4 & \nu_3 \bnu_3  & \nu_1 \bnu_3 & \bnu_1 \nu_3 \\
-\nu_2 \bnu_{2} & - \bnu_2 \nu_3 - \nu_1 \bnu_4 & 0 & \nu_4 \bnu_4  & -\bnu_2 \nu_4 & -\nu_2 \bnu_4 \\
\bnu_1 \nu_{4} -\bnu_2 \nu_{3}  & -\nu_3 \bnu_3 & -\nu_4 \bnu_4 & 0  & -\bnu_3 \nu_4 & -\nu_3 \bnu_4\\
-\nu_1 \bnu_2 &  -\nu_1 \bnu_3 &  \bnu_2 \nu_4 &  \bnu_3 \nu_4 & 0 & \bnu_2 \nu_3 -\nu_2 \bnu_3 \\
-\bnu_1 \nu_2 & -\bnu_1 \nu_3  &  \nu_2 \bnu_4 &  \nu_3 \bnu_4 & \nu_2 \bnu_3 - \bnu_2 \nu_3 & 0
\end{array}
\right)
\end{equation}
with an additional constraint: $\nu_1 \bnu_4 - \bnu_1 \nu_4 = \nu_2 \bnu_3 - \bnu_2 \nu_3$. Here $\nu_i$ and $\bar{\nu}_i$ satisfy already periodic/anti-periodic constraints ($\sigma = \pm 1$):
\begin{eqnarray}
{ \tilde \nu }_i = \sigma_i \nu_i^{[2]} = \sigma \, \nu_i^{[2]}, \label{nuanalytcont}
\end{eqnarray}
where here and in what follows $f^{\left[n\right]}\left(u\right)  =f\left(u+in/2\right)$. To describe anomalous dimensions of $sl(2)$-like states (\ref{sl2op}) it is enough to consider ${\bf P}\mu$-system reduced to symmetric, parity invariant states. The reduced ${\bf P}\mu$-system is identified by constraints $\bP_5 = \bP_6 = \bP_0$, $\nu_i = \bar{\nu}_i$ and is written as \cite{ABJMQSC,ABJMQSCdetailed,ABJMQSC12loops}:
\begin{eqnarray}
{ \tilde \nu}_i &=& - \bP_{ij} \, \chi^{jk} \nu_k  \label{nuPmonodromy} ,\\
{ \widetilde {\bP} }_{ij} - { \bf P }_{ij} &=& \nu_i { \tilde \nu }_j - \nu_j { \tilde \nu }_i , \label{Pdiscont} 
\end{eqnarray}  
where 
\begin{eqnarray}
\bP_{ij} = \left( \begin{array}{cccc} 0 & - \bP_1 & - \bP_2 & - \bP_0 \\  \bP_1 & 0 & - \bP_0 & - \bP_3 \\ \bP_2 & \bP_0 & 0 & - \bP_4 \\ \bP_0 & \bP_3 & \bP_4 & 0 \end{array} \right) , \;\;\;\; \chi^{ij} = \left( \begin{array}{cccc} 0 & 0 & 0& - 1 \\  0 & 0 & 1 & 0 \\ 0 & -1 & 0 & 0 \\ 1 &  0& 0 & 0 \end{array} \right) ,
\end{eqnarray}
and 
\begin{eqnarray}
( \bP_0 )^2 &=& 1 - \bP_1 \bP_4 + \bP_2 \bP_3 . \label{P0}
\end{eqnarray}
In addition to the above constraints it is required \cite{ABJMQSC12loops}, that $\bP$ and $\nu$ functions have no poles and stay bounded at branch points. The quantum numbers of the states we are interested in, that is twist L, spin S and conformal dimension $\triangle$ are encoded in the behavior of $\bP$, $\nu$ functions at large $u$ \cite{ABJMQSC,ABJMQSCdetailed,ABJMQSC12loops}:  
\begin{eqnarray}
\bP_{0-4} &\simeq& (A_0 u^0, \, A_1 u^{-L}, \,A_2 u^{-L-1}, \,A_3 u^{+L+1}, \,A_4 u^{+L}) , \nonumber\\
A_1 A_4 &=&-\frac{(\Delta -L+S) (\Delta -L-S+1) (\Delta +L-S+1) (\Delta +L+S)}{L^2 (2 L+1)} \nonumber\\
A_2 A_3 &=&-\frac{(\Delta -L+S-1) (\Delta -L-S) (\Delta +L-S+2) (\Delta +L+S+1)}{(L+1)^2 (2 L+1)},\label{Pasympt}
\end{eqnarray}
and  
\begin{eqnarray}
\nu_i \sim \left( u^{\Delta - L} ,u^{\Delta+1}, u^{\Delta} , u^{\Delta+L+1}  \right).
\end{eqnarray}
The anomalous dimension $\gamma$ is given by $\gamma = \triangle - L - S$.

\section{Perturbative solution of ABJM QSC}\label{PerturbativeSolution}

For the perturbative solution of ABJM quantum spectral curve we use the same set of equations as in \cite{ABJMQSC12loops}. The latter easily follow\footnote{See \cite{ABJMQSC12loops} for details} from fundamental $\bP \nu$-system  (\ref{nuPmonodromy})-(\ref{Pdiscont}) and are given by
\begin{align}
\frac{\nu_1^{[3]}}{\bP_1^{[1]}} - \frac{\nu_1^{[-1]}}{\bP_1^{[-1]}} - \s \left(
\frac{\bP_0^{[1]}}{\bP_1^{[1]}} - \frac{\bP_0^{[-1]}}{\bP_1^{[-1]}}
\right)\nu_1^{[1]} &= -\s \left(
\frac{\bP_2^{[1]}}{\bP_1^{[1]}} - \frac{\bP_2^{[-1]}}{\bP_1^{[-1]}}
\right)\nu_2^{[1]}\, , \label{Baxternu1} \\
\frac{\nu_2^{[3]}}{\bP_1^{[1]}} - \frac{\nu_2^{[-1]}}{\bP_1^{[-1]}} + \s \left(
\frac{\bP_0^{[1]}}{\bP_1^{[1]}} - \frac{\bP_0^{[-1]}}{\bP_1^{[-1]}}
\right)\nu_2^{[1]} &= \s \left(
\frac{\bP_3^{[1]}}{\bP_1^{[1]}} - \frac{\bP_3^{[-1]}}{\bP_1^{[-1]}}
\right)\nu_1^{[1]}\, , \label{Baxternu2}
\end{align}
and 
\begin{align}
\s \nu_1^{[2]} &= \bP_0 \nu_1 - \bP_2 \nu_2 + \bP_1 \nu_3 \, , \label{nu3sol} \\
\s \nu_2^{[2]} &= -\bP_0 \nu_2 + \bP_3 \nu_1 + \bP_1 \nu_4 \, , \label{nu4sol} \\
\btP_2 - \bP_2 &= \s \left(
\nu_3 \nu_1^{[2]} - \nu_1 \nu_3^{[2]}
\right)\, , \label{pt2} \\
\btP_1 - \bP_1 &= \s \left(
\nu_2 \nu_1^{[2]} - \nu_1 \nu_2^{[2]}
\right)\, , \label{pt1} \\
\left(
\nu_1 + \s \nu_1^{[2]}
\right) \left(
\bp_0 - (h x)^L
\right) &= \bp_2 \left(
\nu_2 + \s \nu_2^{[2]}
\right) - \bp_1 \left(
\nu_3 + \s \nu_3^{[2]}
\right)\, , \label{p0}
\\ 
\left(
\nu_2 + \s \nu_2^{[2]}
\right)\left(
\bp_0 + (h x)^L
\right) &= \bp_3 \left(
\nu_1 + \s \nu_1^{[2]}
\right) + \bp_1 \left(
\nu_4 + \s \nu_4^{[2]}
\right)\, ,  \label{p3}
\end{align}
where 
\begin{equation}
x\equiv x (u) = \frac{u+\sqrt{u^2 - 4 h^2}}{2 h}
\end{equation}
is the Zhukovsky variable parameterizing the single cut of $\bP$ functions on the defining Riemann sheet. In addition from the analytical structure of $\nu_i (u)$ functions on the defining Riemann sheet it follows, that the following combinations of functions
\begin{align}
\nu_i (u) + \tilde{\nu}_i (u) &= \nu_i (u) + \s \nu_i^{[2]} (u)\, , \nonumber \\
\frac{\nu_i (u) - \tilde{\nu}_i (u)}{\sqrt{u^2 - 4 h^2}} &=
\frac{\nu_i (u) - \s \nu_i^{[2]} (u)}{\sqrt{u^2 - 4 h^2}}\, 
\label{cutsfree}
\end{align}
are free of cuts on the whole real axis. Similar to \cite{VolinPerturbativeSolution,ABJMQSC12loops} we will parametrize the $\bP (u)$ functions entering the solution as
\begin{align}\label{eq:P1}
\bP_1 &= (x h)^{-L} \bp_1 = (x h)^{-L} \left(
1+\sum_{k=1}^{\infty}\sum_{l=0}^{\infty} c_{1,k}^{(l)}\frac{h^{2 l+k}}{x^k}
\right)\, , \\
\bP_2 &= (x h)^{-L} \bp_2 = (x h)^{-L}\left(
\frac{h}{x} + \sum_{k=2}^{\infty}\sum_{l=0}^{\infty} c_{2,k}^{(l)} \frac{h^{2 l+k}}{x^k}
\right)\, , \\
\bP_0 &= (x h)^{-L} \bp_0  = (x h)^{-L} \left(
\sum_{l=0}^{\infty} A_0^{(l)} h^{2 l} u^L + 
\sum_{j=0}^{L-1}\sum_{l=0}^{\infty} m_j^{(l)} h^{2 l} u^j
+\sum_{k=1}^{\infty}\sum_{l=0}^{\infty} c_{0,k}^{(l)}
\frac{h^{2 l+k}}{x^k}
\right)\, , \\
\bP_3 &= (x h)^{-L} \bp_3 = (x h)^{-L}\left(
\sum_{l=0}^{\infty} A_3^{(l)} h^{2 l} u^{2 L+1}
+ \sum_{j=0}^{2 L}\sum_{l=0}^{\infty} k_j^{(l)} h^{2 l} u^j
+ \sum_{k=1}^{\infty}\sum_{l=0}^{\infty} c_{3,k}^{(l)}\frac{h^{2 l+k}}{x^k}
\right) \, .\label{eq:P3}
\end{align}
where we have accounted for correct polynomial asymptotic of $\bP$ functions at large values of spectral parameter $u$ (\ref{Pasympt}). We may always assume that on the defining or first Riemann sheet $|x(u)| > 1$ and thus the above expansions are justified. Due to gauge symmetry of QSC equations\footnote{See for details \cite{ABJMQSC12loops}.}
\begin{equation}
\nu_i\to R_i^{~j} \nu_j , \quad \bP_{ij}\to R_i^{~i'} \bP_{i'j'} R_j^{~j'}\, ,
\end{equation}
where $R$ is any $4\times 4$ constant matrix satisfying $R^t\chi R = \chi$ the coefficients $m_j^{(l)}$, $k_j^{(l)}$ in the above parametrization at twist $L=1$ are left undetermined. The coefficients $A_0^{(l)}$, $A_3^{(l)}$ and $c_{i,k}^{(l)}$ are some functions of spin $S$ only, otherwise they are just constants. Here we have also used the mentioned gauge freedom to set $A_1 = 1$ and $A_2 = h^2$.
Since $x$ tends to $x_+(u)=\frac{u + i\sqrt{4h^2-u^2}}{2h}$ and $x_-(u)=1/x_+(u)=\frac{u - i\sqrt{4h^2-u^2}}{2h}$ on the upper and lower bank of the cut, correspondingly, the values of the functions $\bP_a(u)$ on the two banks are related as 
$$\bP_a(u-i0)=\left.\bP_a(u+i0)\right|_{x_+(u)\to 1/x_+(u)}\,.$$ 
Therefore, in a sufficiently small vicinity\footnote{The vicinity is such that the substitution $x\to 1/x$ retains the convergence of \eqref{eq:P1}-\eqref{eq:P3}} of the cut on the second sheet of $u$-plane (the inner vicinity of unit circle in $x$-plane) we have
\begin{equation}
\btP_a =\bP_a\Big|_{x\to 1/x} = \left(\frac{x}{h}\right)^L \btp_a\, , \quad \btp_a = \bp_a\Big|_{x\to 1/x}\, .
\end{equation}

Next, the expansion of $\nu_i (u)$ functions in terms of QSC coupling constant $h$ is given by
\begin{equation}
\nu_{i} (u) = \sum_{l=0}^{\infty} h^{2 l - L} \nu_i^{(l)} (u)\,.
\end{equation}

\subsection{Leading order}
From now on we consider the case of $L=1$ operators. First, from Eqs. \eqref{Pasympt} and \eqref{P0} at large values of spectral parameter $u$ we get
\begin{equation}
A_0^{(0)}=\sigma (2S+1)\,.
\end{equation} 
Next, we take LO approximation of the first Baxter equation (\ref{Baxternu1}). The solution of the latter, as described in detail in the next section, is given by
\begin{equation}
\nu_1^{(0)} (u) = \alpha Q^{[-1]} (S, u)\, ,
\end{equation} 
where $Q (S, u)$ is LO Baxter polynomial defined in Eq. \eqref{eq:BaxterPolynomial} and $\alpha$ is some spin $S$ dependent constant to be determined later. In general for the details of the solutions of Baxter equations (\ref{Baxternu1})-(\ref{Baxternu2}) we refer the reader to  next section. Here we will just use the results obtained there. Next, from equation (\ref{nu3sol}) we determine the expression for $\nu_3^{(0)} (u)$ and substitute it in the equation (\ref{pt2}). Expanding the latter at $u=0$ up to $\OO (u^2)$ we get the expression\footnote{See Appendix \ref{BaxterDerivatives} for definition of $B$-sums.}  for constant $\alpha$:
\begin{equation}
\frac{1}{\alpha^2} = -4 i B_1 (S)\, ,\quad B_1 (S) = H_1 (S) - H_{-1} (S)\, . \label{alpha}
\end{equation}
Also from the requirement of absence of poles in combinations (\ref{cutsfree}) for $\nu_1^{(0)}$ we may determine the value of $\s = (-1)^S$.  Knowing the expression for $\nu_1^{(0)} (u)$ we may determine $\nu_2^{(0)}$ by solving second Baxter equation (\ref{Baxternu2}): 
\begin{multline}
\nu_{2}^{\left(0\right)}  =\sigma\frac{\alpha}{2}\bigg[-\frac{1}{4}A_3^{(0)}\bigg(\frac{3\left(S+2\right)\left(S+1\right)}{2S+3}\,Q^{[-1]}\left(S+2,u\right)-2\frac{3S^{2}+3S+1}{2S+1}Q^{[-1]}\left(S,u\right) \\ +\frac{3S\,\left(S-1\right)}{2S-1}Q^{[-1]}\left(S-2,u\right)\bigg)+\frac{i}{2}k_2^{(0)}\,\left[Q^{[-1]}\left(S+1,u\right)-\delta_{S\neq0}Q^{[-1]}\left(S-1,u\right)\right] \\ +k_1^{(0)}\frac{1}{2S+1}\,Q^{[-1]}\left(S,u\right)\bigg]\,.
\end{multline}
Next, from equation (\ref{pt1}) expanded at  $u=0$ up to $\OO (u)$ we get the value of $A_3^{(0)}$ constant:
\begin{equation}
 A_3^{(0)} = -\frac{4}{3} (2S+3)(2S-1) B_1 (S)\,.
\end{equation} 
In addition, from the same expansion we get the following values of coefficients
\begin{equation}
k_2^{(0)} = 0\, , \quad c_{1,1}^{(0)} = 0\,.
\end{equation}
Note, that if we account for $\delta_{S,0}$ term in the $\nu_2^{(0)}$ solution then the value of $k_2^{(0)}$ coefficient turns out to be fixed. So, the analytical continuation in spin $S$ allows us to fix extra gauge freedom.

\subsection{Next-to-leading order}

Before starting actual NLO calculation it makes sense to determine as many required constants as possible with the information on LO solutions we already have. Performing small $u$ expansion of (\ref{pt1}) up to $\OO (u^3)$ we get
\begin{equation}
c_{1,2}^{(0)} = 4B_1(S) -B_1(S)^2-2B_2(S)
\, .
\end{equation}
Next, equation (\ref{p0}) expanded at $u=0$ up to $\OO (1)$ terms gives us the value of $c_{0,1}^{(0)}$ coefficient:
\begin{equation}
c_{0,1}^{(0)} = \frac{\s k_1^{(0)}}{2(1 +2 S)} - 1 - \frac{i \s S\, (1+S)}{3 (1+ 2 S) \alpha^2}\, .
\end{equation}
Substituting the expression for $\nu_4^{(0)}$ from (\ref{nu4sol}) into equation (\ref{p3}) and expanding the latter at small $u$ up to $\OO (1)$ terms we get the value of $c_{3,1}^{(0)}$ coefficient:
\begin{equation}
c_{3,1}^{(0)} = \frac{\s (2 S (1+S)+3 i \alpha^2 k_1^{(0)}) (12 i (1 + 2 S)\alpha^2 - \s (2 S (1+S) + 3 i \alpha^2 k_1^{(0)}) )}{36 (1+ 2 S)^2\alpha^4}\, .
\end{equation}
Finally, the expansion of (\ref{pt2}) at $u=0$ up to $\OO (u^4)$ terms gives us the values of $c_{2,2}^{(0)}$ and $c_{2,3}^{(0)}$ coefficients:
\begin{equation}
c_{2,2}^{(0)} = 0\, ,\quad c_{2,3}^{(0)} = 4 i\alpha^2 \left(
B_1 (S) B_2 (S) + B_3 (S)
\right) \, .
\end{equation}

Now, we are ready to proceed with the solution of NLO Baxter equations. As is explained in the next section the solution of the first Baxter equation (\ref{Baxternu1}) at NLO is given by
\begin{multline}
\nu_1^{(1)} (u) = \frac{\alpha}{2} \left(
\s A_0^{(1)} - 2 (2 S - 1) B_1 (S)
\right) 
 \Big\{ 
Q^{[-1]} (S, u) \left(
\egamma + \log2 - i\eta_1 (u) \right. \\ \left. - H_1 (S) + i \pi \coth (\pi u)
\right) + \sum_{k=1}^S \frac{1+(-1)^k}{k} Q^{[-1]} (S-k, u)
\Big\} \\
- \alpha \sum_{k=1}^S \frac{1+(-1)^k}{k} \left(
B_1 (S) {+} B_1 (S-k)
\right) Q^{[-1]} (S-k, u) \\ + \phi_{1,0}^{\mathrm{per}} Q^{[-1]} (S, u)
+ \phi_{1,1}^{\mathrm{per}} \P_1 (u) Q^{[-1]} (S, u)\,.
\end{multline}
The absence  of poles in combinations (\ref{cutsfree}) for $\nu_1^{(1)}$ allows us to determine the values of coefficients $A_0^{(1)}$ and $\phi_{1,1}^{\mathrm{per}}$:
\begin{equation}
A_0^{(1)} = 2 \sigma\, (3 + 2 S) B_1 (S)\, ,\quad \phi_{1,1}^{\mathrm{per}} = -2 i\alpha B_1 (S)\, .
\end{equation}
Next, the expansions of (\ref{pt2}) at $u=0$ up to $\OO (u^2)$ terms fixes the value of $\phi_{1,0}^{\mathrm{per}}$:
\begin{equation}
\phi_{1,0}^{\mathrm{per}} = \alpha\Big\{
\frac43 B_1(S)^2+B_2(S)+\frac{3B_3(S)+2H_3(S)-2H_{-3}(S)}{3B_1(S)}- 2 B_1 (S) (1+2\log2)\Big\}\,.
\end{equation}

Now we are ready to solve the second Baxter equation at NLO (\ref{Baxternu2}). The details of the solution could be found in the next section. In terms of $q_2^{(1)} (u)$ (\ref{q21}) the expression for $\nu_2^{(1)} (u)$ is then given by
\begin{equation}
\nu_2^{(1)} (u) = q_2^{(1)} (u-i/2)\,.
\end{equation} 
Requiring the absence of poles in combinations (\ref{cutsfree}) for $\nu_2^{(1)}$ allows us to fix coefficients $ \phi_{2,0}^{\mathrm{per}}$ and $\phi_{2,0}^{anti}$ :
\begin{equation}
\phi_{2,0}^{\mathrm{per}} = 0\, ,\quad  \phi_{2,0}^{anti} = 4 i\alpha B_1 (S)\, .
\end{equation}
Finally, expanding equation (\ref{pt1}) at $u=0$ up to $\OO (1)$ terms gives us the value of $A_3^{(1)}$ coefficient:
\begin{multline}
A_3^{(1)} = 
-\frac{16}{3} (2 S-1) (2 S+3)\left(3 \bar{H}_{-2,-1}-2
\bar{H}_{-2,i}-\bar{H}_{-2,1}-\bar{H}_{-1,-2
}+2 \bar{H}_{-1,2 i}-\bar{H}_{-1,2}
\right.
\\
-6
\bar{H}_{i,-2}+12 \bar{H}_{i,2 i}-6
\bar{H}_{i,2}-6 \bar{H}_{2 i,-1}+4
\bar{H}_{2 i,i}+2 \bar{H}_{2
	i,1}-\bar{H}_{1,-2}+2 \bar{H}_{1,2
	i}-\bar{H}_{1,2}+3 \bar{H}_{2,-1}
\\
-2
\bar{H}_{2,i}-\bar{H}_{2,1}+2
\bar{H}_{-1,i,-1}-2 \bar{H}_{-1,i,1}+8
\bar{H}_{i,-1,-1}-12 \bar{H}_{i,-1,i}+4
\bar{H}_{i,-1,1}-16 \bar{H}_{i,i,-1}\\
\left.
+16
\bar{H}_{i,i,i}+4 \bar{H}_{i,1,-1}-4
\bar{H}_{i,1,i}+2 \bar{H}_{1,i,-1}-2
\bar{H}_{1,i,1}
-\frac12 B_1 \zeta_2
\right)- \frac{4}{3} (5 + 20 S + 4 S^2) B_1^2\,.
\end{multline}
where $\bar{H}_{a,\,\ldots}= H_{a,\,\ldots}(2S)$ is defined in  \eqref{eq:sums}. The reason behind the appearance of sums different from harmonic is related to the fact that solutions of Baxter equations at NLO contain Baxter polynomials $Q(S, u)$ under different summation signs with various weights. For example, in the case of homogeneous solution of second Baxter equation  $\Z (S,u)$ entering the expression for $q_2^{(1)}(u)$ \eqref{q21} its expansion at 
$u = i/2$ is given by
\begin{equation}
\Z (S, u + \frac{i}{2}) = -i (-1)^S [S_{-1} (S) +\ln 2] + u(-1)^S\left[ - B_1 (S)\ln2
+ V(S) + {\zeta_2}/{2} \right] + \OO (u^2)\,,
\end{equation} 
where $ V(S)  = \sum_{k=1}^S\frac{1+(-1)^{S+k}}{k+S} B_1 (k-1)$ could be further rewritten in terms of generalized harmonic sums \eqref{eq:sums}. 
Also, the necessity of the argument $2S$ could be promptly realized  after examining the denominators of the rational numbers entering the results for large enough $S$ and observing the appearance of prime numbers in the interval $[S+1,2S)$ among their factors.

%
We would like to mention, that in the simplification of the coefficients expressions the use of HarmonicSums mathematica package \cite{Ablinger1,Ablinger2,Ablinger3,Ablinger4,BlumleinStructuralRelations,RemiddiVermaseren,Vermaseren} was helpful.

\section{Solution of  Baxter equations with Mellin transform}\label{BaxterEquationsMellin}

In the previous section we have seen that the most complicated part of the QSC solution is the solution of two inhomogeneous Baxter equations at each perturbation order. To solve these second order finite difference equations we will employ Mellin transform technique to convert them to ordinary differential equations. The latter was originally applied to the solution of Lipatov's reggeon spin chain \cite{LipatovSpinChain} by Faddeev and Korchemsky in \cite{FaddeevKorchemsky}. Later this technique was used to solve asymptotic Baxter equation in $\mathcal{N}=4$ SYM up to three and four loops \cite{Kotikov1,Kotikov2}. In order to iteratively search for the perturbative solution of equations \eqref{Baxternu1}--\eqref{p3}, we expand Eqs. \eqref{Baxternu1}, \eqref{Baxternu2} up to $h^k$ and obtain the inhomogeneous equations for $q_{1,2}^{(k)}=\left(\nu_{1,2}^{(k)}\right)^{[1]}$ in the following form
\begin{align}
(u+i/2)q_{1}^{(k)}(u+i)-i(2S+1)q_{1}^{(k)}(u)-(u-i/2)q_{1}^{(k)}(u-i)=&V_{1}^{(k)} \label{Baxter1a}\,,\\
(u+i/2)q_{2}^{(k)}(u+i)+ i(2S+1)q_{2}^{(k)}(u)-(u-i/2)q_{2}^{(k)}(u-i)=&V_{2}^{(k)}\,. \label{Baxter2a}
\end{align}
Here $V_{1}^{(k)}$ depends on $q_{1,2}^{(l)}$ with $l<k$, and $V_{2}^{(k)}$ depends in addition on $q_1^{(k)}$.
Applying Mellin transformation\footnote{See Appendix \ref{MellinTransformation} for more details and notation.} to Eqs. \eqref{Baxter1a} and \eqref{Baxter2a}, we get
\begin{equation}
\left(\bar{z}-z\right)\partial_z\Psi_{1}^{(k)}(z)+2S\Psi_{1}^{(k)}(z)=i\widetilde{V}_{1}^{(k)}\left(z\right)\,, \label{Baxter1Mellin}
\end{equation}
\begin{equation}
\left(\bar{z}-z\right)\partial_z\Psi_{2}^{(k)}(z)-2\left(S+1\right)\Psi_{2}^{(k)}(z)=i\widetilde{V}_{2}^{(k)}\left(z\right)\,, \label{Baxter2Mellin}
\end{equation}  
where $\Psi_{1,2}^{(k)}=\mathcal{M}^{-1}[q_{1,2}^{(k)}]$ and $\widetilde{V}_{1,2}^{(k)}=\mathcal{M}^{-1}[V_{1,2}^{(k)}]$. The integration of equations (\ref{Baxter1Mellin}) and (\ref{Baxter2Mellin}) is straightforward and is given by 
\begin{gather}
\Psi_{1}^{(k)}(z)=i\left(\bar{z}-z\right)^{S}\int\left(\bar{z}-z\right)^{-S-1}\widetilde{V}_{1}^{(k)}\left(z\right)dz\,,\label{Baxter1SolutionMellin}\\
\Psi_{2}^{(k)}(z)=i\left(\bar{z}-z\right)^{-S-1}\int\left(\bar{z}-z\right)^{S}\widetilde{V}_{2}^{(k)}\left(z\right)dz\,.\label{Baxter2SolutionMellin}
\end{gather}

Note that, when passing to the Mellin space, we silently assumed that $\Psi_{1,2}^{(k)}(z)$ are finite at $z=0$ and $z=1$. In general it might be not so due to the appearing logarithms of $z$ and $\bar z$. Therefore, the approach based on Mellin transformation, should be used with great care, in particular the results obtained within this approach should be transformed back to $u$-space and directly checked against the equations \eqref{Baxter1a} and \eqref{Baxter2a}. These complications may be viewed as disadvantages of the Mellin-space approach. Nevertheless, we find it advantageous to use Mellin transformation technique, at least, up to the next-to-leading order considered in this paper.

\subsection{Homogeneous solution}

The solution of homogeneous first Baxter equation (\ref{Baxter1Mellin}) in Mellin space is easy and is given by\footnote{The arbitrary constant in front of solution is dropped.} 
\begin{equation}
\Psi_{1}^{\left(0\right)} = \left(\bar{z}-z\right)^{S}\,.
\end{equation}
Its Mellin transform to spectral parameter $u$-space is given by
\begin{multline}\label{eq:BaxterPolynomial}
Q\left(S,u\right) =\,_{2}F_{1}\left(-S,\frac{1}{2}+iu;1;2\right)\\=\frac{(-1)^{S}\Gamma\left(\frac{1}{2}+iu\right)}{S!\Gamma\left(\frac{1}{2}+iu-S\right)}\,_{2}F_{1}\left(-S,\frac{1}{2}+iu;\frac{1}{2}+iu-S;-1\right)\,.
\end{multline} 
Moreover, $\Phi_Q^{per} (u) Q (S,u)$ and $\Phi_Q^{anti} (u) Q (S,u)$, where $\Phi_Q^{per} (u)$ and $\Phi_Q^{anti} (u)$ are arbitrary periodic and anti-periodic functions of spectral parameter $u$, are also solutions of homogeneous first (\ref{Baxter1a}) and second (\ref{Baxter2a}) Baxter equations correspondingly. 

To find second solutions of homogeneous Baxter equations let us consider second Baxter equation (\ref{Baxter2a}). Making the ansatz $q_2(S,u) = Q(S,u) b^{[1]}(S,u)$ similar to \cite{VolinPerturbativeSolution,ABJMQSC12loops} the homogeneous Baxter equation (\ref{Baxter2a}) could be rewritten as 
\begin{equation}
\nabla_{+} (u Q^{[1]} Q^{[-1]}\nabla_{-}b) = 0\, , \label{Baxter2rewritten}
\end{equation}
where $\nabla_{+} f = f - f^{[2]}$ and $\nabla_{-} f = f + f^{[2]}$. From equation (\ref{Baxter2rewritten}) it follows then, that\footnote{Of course, $b(u)$ is defined up to arbitrary multiplicative periodic and additive anti-periodic constants which is taken into account in Eqs. \eqref{eq:homo1}, \eqref{eq:homo2}.}
\begin{equation}
b(u)+b(u+i)=\frac{1}{uQ^{[1]}Q^{[-1]}}\,.
\end{equation}
To solve this difference equation we will use the empirically guessed identity
\begin{multline}
\frac{1}{uQ^{[1]}Q^{[-1]}}=\frac{(-1)^{S}}{u}+i(-1)^{S}\sum_{k=0}^{\left\lfloor \frac{S-1}{2}\right\rfloor }\frac{1}{S-k}\\
\times\left(\frac{Q\left(S-1-2k,u-\frac{i}{2}\right)}{Q\left(S,u-\frac{i}{2}\right)}+\frac{Q\left(S-1-2k,u+\frac{i}{2}\right)}{Q\left(S,u+\frac{i}{2}\right)}\right)\,.
\end{multline}
Then we see, that
\begin{align}
b(u)+b(u+i) 
& =\frac{(-1)^{S}}{u}+\left[i(-1)^{S}\sum_{k=0}^{\left\lfloor \frac{S-1}{2}\right\rfloor }\frac{1}{S-k}\frac{Q\left(S-1-2k,u-\frac{i}{2}\right)}{Q\left(S,u-\frac{i}{2}\right)}+\left(u\to u+i\right)\right]\, .
\end{align}
Introducing Hurwitz function $\eta_{-1} (u)$ defined as 
\begin{equation}
\eta_{-1}(u)=\sum_{n=0}^{\infty}\frac{\left(-1\right)^{n}}{u+in}=\frac{i}{2}\left(\psi\left(-i\frac{u}{2}\right)-\psi\left(-i\frac{u+i}{2}\right)\right)\,, 
\end{equation}
where $\psi (u)$ is polygamma function and noting, that 
\begin{equation}
\eta_{-1}(u)+\eta_{-1}(u+i)=\frac{1}{u}\,,
\end{equation}
we see that the solution for $b(u)$ function is given by
\begin{equation}
b(u)=(-1)^{S}\eta_{-1}(u)+i(-1)^{S}\sum_{k=0}^{\left\lfloor \frac{S-1}{2}\right\rfloor }\frac{1}{S-k}\frac{Q\left(S-1-2k,u-\frac{i}{2}\right)}{Q\left(S,u-\frac{i}{2}\right)}\, .
\end{equation}
Finally the expression for second solution with polynomial asymptotic, which we will denote by $\mathcal{Z}(S,u)$ is given by\footnote{For a rigorous proof that $\Z (S,u)$ is a solution see Appendix \ref{BaxterDerivatives}.}
\begin{equation}
\Z (S,u) = i\sigma\sum_{k=0}^{\left\lfloor \frac{S-1}{2}\right\rfloor }\frac{1}{S-k}Q\left(S-1-2k,u\right)+\sigma\eta_{-1}(u+{i}/{2})Q\left(S,u\right)\,.
\end{equation}
Once this solution is found, one can check directly that it satisfies the homogeneous part of Eq. \eqref{Baxter2a}, using the generating function found in Appendix \ref{BaxterDerivatives}.

The general solutions of first and second homogeneous Baxter equations are then given by
\begin{align}\label{eq:homo1}
q_1^{hom} (S,u) &= \Phi_{1,per} Q(S,u) + \Phi_{1,anti} \Z (S,u)\, , \\
\label{eq:homo2}
q_2^{hom} (S,u) &= \Phi_{2,anti} Q(S,u) + \Phi_{2,per} \Z (S,u)\, ,
\end{align}
where $\Phi_{i,per}$ and $\Phi_{i,per}$  are arbitrary periodic and anti-periodic functions in spectral parameter $u$. Otherwise they are arbitrary functions of spin $S$ to be determined from consistency conditions as described in previous section. We will parametrize their $u$ dependence similar to \cite{VolinPerturbativeSolution,ABJMQSC12loops} with the basis of periodic and anti-periodic combinations of Hurwitz functions defined as 
\begin{equation}
\P_k (u) = \eta_k (u) + \sign(k)(-1)^k\eta_k (i-u) = \text{sgn}(k)\P_k (u+i)\, , \quad k\neq 0 \in \mathbb{Z}\, ,
\end{equation}
where 
\begin{equation}
\eta_a  (u) = \sum_{k=0}^{\infty}\frac{(\text{sgn}(a))^k}{(u + i k)^{|a|}}\, .
\end{equation}
Note that $\P_k (u)$ can be expressed via elementary functions:
\begin{equation}
\P_k (u) = \frac{(-\partial_u)^{|k|-1}}{(|k|-1)!} \begin{cases} 
\pi\coth(\pi u) & k > 0 \\
\pi/\sinh(\pi u)  & k < 0
\end{cases}\,.
\end{equation}

Then the functions  $\Phi_{a}^{\mathrm{per}}$ and $\Phi_{a}^{\mathrm{anti}}$ are written as
\begin{equation}
\Phi_{a}^{\mathrm{per}} (u) = \phi_{a,0}^{\mathrm{per}} + \sum_{j=1}^{\Lambda}\phi_{a,j}^{\mathrm{per}} \P_j (u)\, , \quad \Phi_{a}^{\mathrm{anti}} (u) =  \sum_{j=1}^{\Lambda}\phi_{a,j}^{\mathrm{anti}}\P_{-j} (u)\,,
\end{equation}
where $\Lambda$ is a cutoff dependent on the order of perturbation theory.

\subsection{Inhomogeneous solution}

Let us now proceed with the solution of inhomogeneous Baxter equations. 

\subsubsection{LO}

At leading order $V_1^{(0)} = 0$ and LO order solution of the first Baxter equation is given by the solution of homogeneous equation:
\begin{equation}
\Psi_1^{(0)} = \alpha (\bar z - z)^S\, ,\quad  q_{1}^{\left(0\right)}(u) = \alpha Q (S,u)\,,
\end{equation}
with $\alpha$ given by equation (\ref{alpha}). The leading order second Baxter equation is already inhomogeneous with $V_2^{(0)}$, after substitution of the anzats for $\bP$-functions, given by 
\begin{align}
V_2^{(0)} &= i\sigma\left[A_{3}^{\left(0\right)}\left(3 u^{2}-\frac{1}{4}\right)+2k_{2}^{\left(0\right)}u+k_{1}^{\left(0\right)}\right] q_1^{(0)}\,, \\
\widetilde{V}_{2}(z) &= i\sigma\left[A_{3}^{\left(0\right)}\left(3\hat{u}^{2}-\frac{1}{4}\right)+2k_{2}^{\left(0\right)}\hat{u}+k_{1}^{\left(0\right)}\right] \Psi_1^{(0)}\,, 
\end{align}  
and the particular solution for $\Psi_2^{(0)}$ is given by (\ref{Baxter2SolutionMellin}) : 
\begin{align}
\Psi_{2}^{\left(0\right)}(z) & =-\sigma\left(\bar{z}-z\right)^{-S-1}\int\left(\bar{z}-z\right)^{S}\left[A_{3}^{\left(0\right)}\left(3\hat{u}^{2}-\frac{1}{4}\right)+2k_{2}^{\left(0\right)}\hat{u}+k_{1}^{\left(0\right)}\right]\Psi_{1}^{\left(0\right)}dz \nonumber \\
& =-\sigma\alpha\left(\bar{z}-z\right)^{-S-1}\int\left(\bar{z}-z\right)^{S}\left[A_{3}^{\left(0\right)}\left(3\hat{u}^{2}-\frac{1}{4}\right)+2k_{2}^{\left(0\right)}\hat{u}+k_{1}^{\left(0\right)}\right]\left(\bar{z}-z\right)^{S}dz\,.
\end{align}
Acting by $\hat{u}$ operator (see Eq. \eqref{eq:hat_u}), we obtain
\begin{multline}
\Psi_{2}^{\left(0\right)}(z)  =\sigma\frac{1}{2}\alpha\left(\bar{z}-z\right)^{-S-1}\int d\left(\bar{z}-z\right)\,\left(\bar{z}-z\right)^{S}\bigg[-\frac{1}{4}A_{3}^{\left(0\right)}\left(3\left(S+2\right)\left(S+1\right)\left(\bar{z}-z\right)^{S+2} \right. \\ \left. -2\left[3S^{2}+3S+1\right]\left(\bar{z}-z\right)^{S}+3S\left(S-1\right)\left(\bar{z}-z\right)^{S-2}\right)\\
 +ik_{2}^{\left(0\right)}\left[\left(S+1\right)\left(\bar{z}-z\right)^{S+1}-S\left(\bar{z}-z\right)^{S-1}\right]+k_{1}^{\left(0\right)}\left(\bar{z}-z\right)^{S}\bigg]\\
 =\sigma\frac{1}{2}\alpha\bigg[C\left(\bar{z}-z\right)^{-S-1}-\frac{1}{4}A_{3}^{\left(0\right)}\Big(3\left(S+2\right)\left(S+1\right)\frac{\left(\bar{z}-z\right)^{S+2}}{2S+3}-2\left[3S^{2}+3S+1\right]\frac{\left(\bar{z}-z\right)^{S}}{2S+1}  \\  +3S\left(S-1\right)\frac{\left(\bar{z}-z\right)^{S-2}}{2S-1}\Big)
 +\frac{i}{2}k_{2}^{\left(0\right)}\left[\left(\bar{z}-z\right)^{S+1}-\left(\bar{z}-z\right)^{S-1}\right]+k_{1}^{\left(0\right)}\frac{\left(\bar{z}-z\right)^{S}}{2S+1}\bigg]\,.
\end{multline}
The requirement that $\Psi_{2}^{\left(0\right)}(z)$ is polynomial fixes constant $C=\frac{i}{2}k_{2}^{\left(0\right)}\delta_{S,0}$ and we finally get\footnote{The homogeneous piece at this order is zero as could be seen from the consistency constraints considered in previous section.}
\begin{multline}
q_{2}^{\left(0\right)}  =\sigma\frac{\alpha}{2}\bigg[-\frac{1}{4}A_{3}^{\left(0\right)}\bigg(\frac{3\left(S+2\right)\left(S+1\right)}{2S+3}\,Q\left(S+2,u\right)-2\frac{3S^{2}+3S+1}{2S+1}Q\left(S,u\right) \\ +\frac{3S\left(S-1\right)}{2S-1}Q\left(S-2,u\right)\bigg)+\frac{i}{2}k_{2}^{\left(0\right)}\left[Q\left(S+1,u\right)-\delta_{S\neq0}Q\left(S-1,u\right)\right]+k_{1}^{\left(0\right)}\frac{1}{2S+1}\,Q\left(S,u\right)\bigg]\,.
\end{multline} 

\subsubsection{NLO}

At NLO the inhomogeneous part of the first Baxter equation after substitution of the anzats for $\bP$-functions  is given by 
\begin{multline}
V_{1}^{(1)}=\frac{4i\sigma q_{2}^{(0)}(u)}{1+4u^{2}}+\left(c_{1,1}^{(0)}+\frac{2}{2u+i}\right)q_{1}^{(0)}(u+i)-\left(c_{1,1}^{(0)}+\frac{2}{2u-i}\right)q_{1}^{(0)}(u-i) \\ +\sigma\left\{ iA_{0}^{(1)}-\frac{4i}{1+4u^{2}}(c_{0,1}^{(0)}-c_{1,1}^{(0)}m_{0}^{(0)})\right\} q_{1}^{(0)}(u)\,.
\end{multline}
To convert it to Mellin space we need the following expressions 
\begin{multline}
\widetilde{\frac{q_{1}^{(0)}(u+i)}{u+i/2}}=\alpha\left[i\bar{z}\partial_{z}z\right]^{-1}\left[-\frac{\bar{z}}{z}\left(\bar{z}-z\right)^{S}{+\frac{1}{z}}\right]=i\frac{\alpha}{z}\partial_{z}^{-1}z^{-1}\left[\left(\bar{z}-z\right)^{S}-\frac{1}{\bar{z}}\right] \\ =i\frac{\alpha}{z}\intop_{0}^{z}\frac{dx}{x}\,\left[\left(\bar{x}-x\right)^{S}-\bar{x}^{-1}\right]=i\frac{\alpha}{z}\left[\ln\bar{z}+G\left(S,z\right)\right]\, ,
\end{multline}
\begin{multline}
\widetilde{\frac{q_{1}^{(0)}(u-i)}{u-i/2}}=\alpha\left[iz\partial_{z}\bar{z}\right]^{-1}\left[-\frac{z}{\bar{z}}\left(\bar{z}-z\right)^{S}{+\frac{\sigma}{\bar{z}}}\right]={-}i\alpha\bar{z}^{-1}\partial_{\bar{z}}^{-1}\left[\frac{1}{\bar{z}}\left(\bar{z}-z\right)^{S}-\frac{\sigma}{z\bar{z}}\right] \\ ={-}i\sigma\frac{\alpha}{\bar{z}}\left[\ln z+G\left(S,\bar{z}\right)\right]\, ,
\end{multline}
where
\begin{align}
G\left(S,z\right) & =\intop_{0}^{z}\frac{dx}{x}\,\left[\left(\bar{x}-x\right)^{S}-\bar{x}^{-1}\right]-\ln\bar{z}=\intop_{0}^{z}\frac{dx}{x}\,\left[\left(\bar{x}-x\right)^{S}-1\right]=\partial_{z}^{-1}z^{-1}\left[\left(\bar{z}-z\right)^{S}-1\right]\,.
\end{align}
The introduced $G$-function satisfies the following recurrence relation
\begin{equation}
G\left(S,z\right)-G\left(S-1,z\right)  =-2\intop_{0}^{z}dx\,\left(1-2x\right)^{S-1}=\frac{1}{S}\left(\left(\bar{z}-z\right)^{S}-1\right)\, , \quad
G\left(0,z\right)  = 0\,.
\end{equation} 
So, we may write it as 
\begin{equation}
G\left(S,z\right)=\sum_{j=1}^{S}\frac{1}{j}\left(\left(\bar{z}-z\right)^{j}-1\right)\,.
\end{equation}
The contribution of the term proportional to $q_{2}^{(0)}(u)$ is determined using partial fractioning 
\begin{equation}
\frac{4i}{1+4u^{2}}=\frac{1}{u-i/2}-\frac{1}{u+i/2}\, ,
\end{equation}  
so that
\begin{multline}
\widetilde{\frac{4iQ\left(S,u\right)}{1+4u^{2}}}  = -i\left[\left(-1+\bar{z}\partial_{z}\right)^{-1}z^{-1}-\left(1+z\partial_{z}\right)^{-1}\bar{z}^{-1}\right]\left(\bar{z}-z\right)^{S}\\
 = -i\left\{ \bar{z}^{-1}G\left(S,z\right)+\sigma z^{-1}G\left(S,\bar{z}\right)+\bar{z}^{-1}\ln z+\sigma z^{-1}\ln\bar{z}+\underline{\left(\bar{z}^{-1}+z^{-1}\sigma\right)B_{1}\left(S\right)}\right\}\,. 
\end{multline}
The underlined term is integration constant chosen in order to get rid of singularities at $z=0$ and $z=1$. Using the above expressions together with the values of constants known at this stage we get
\begin{multline}
\frac{\widetilde{V_{1}^{(1)}}\left(z\right)}{\alpha}  =\frac{i}{2}\left[\left(2S-1\right)\left(\bar{z}-z\right)^{S+2}-\left(2S+3\right)\left(\bar{z}-z\right)^{S} (\bar z -z)(\s -1) + (1+\s)\right]\frac{B_{1}\left(S\right)}{\left(z\bar{z}\right)} \\ +\frac{i}{2}\left(z\bar{z}\right)^{-1}\left((1+\s)(\bar{z} - z) + (1-\s)\right)\delta G\left(S,z\right)+i\sigma A_{0}^{(1)}\left(\bar{z}-z\right)^{S}\, ,
\end{multline}
where
\begin{equation}
\delta G(S, z) \equiv G(S,z) - G(S,\bar z) = \sum_{j=1}^S\frac{1-(-1)^j}{j}(\bar z - z)^j\,.
\end{equation}
Rewriting the above expression as
\begin{align}
\frac{1}{z\bar z}\delta G (S,z) &= H(S,z) - H(S,\bar z) + \left(\frac{1}{z}-\frac{1}{\bar z}\right) B_1 (S)\, , \\
\frac{\bar z -z}{z\bar z}\delta G (S, z) &= H(S,z) + H(S,\bar z) + \left(\frac{1}{z}+\frac{1}{\bar z}\right) B_1 (S)\,,
\end{align}
where
\begin{equation}
H (S, z) = \sum_{j=1}^S \frac{1-(-1)^j}{j z} \left(
(\bar z - z)^j - 1
\right) = -2 \sum_{k=0}^{S-1} (\bar z - z)^k \left\{
B_1 (S) - B_1 (k)
\right\}\,,
\end{equation}
we may get rid from  $z, \bar z$ in the denominator and the final expression for $\widetilde{V}_{1}^{(1)}\left(z\right)$  takes the form  ($w = \bar z -z$):
\begin{equation}
\frac{\widetilde{V}_{1}^{(1)}\left(z\right)}{\alpha i} = w^S (\sigma A_0^{(1)} - 2 (2 S-1) B_1 (S)) + 2\sum_{k=1}^{S}(1+(-1)^{k})
\left\{
B_1 (S) + B_1 (S-k)
\right\} w^{S-k}
\, ,
\end{equation}
Then the particular solution is given by (\ref{Baxter1SolutionMellin}) :
\begin{multline}
\Psi_{1}^{(1)}(z) = \frac{\alpha}{2}\left(
\sigma A_0^{(1)} - 2 (2 S-1) B_1 (S)
\right) w^S\log w \\
-\alpha\sum_{k=1}^S\frac{1+(-1)^k}{k}w^{S-k}\left\{
B_1 (S) + B_1 (S-k)
\right\}\, ,
\end{multline}
For the Mellin transformed\footnote{See the details of transformation in Appendix \ref{MellinForFunctions}} to $u$-space particular solution plus homogeneous piece with at maximum first order poles in $u$ as required by QSC analyticity constraints we then get\footnote{Here we have dropped the piece with non-polynomial asymptotic in the particular solution contribution. The latter is defined up to a homogeneous solution and we may use this freedom to ensure that the ansatz for overall solution has correct polynomial asymptotics we are looking for.}
\begin{multline}
q_1^{(1)} (u) = \frac{\alpha}{2} \left(
\s A_0^{(1)} - 2 (2 S - 1) B_1 (S)
\right) 
\Big\{ 
Q (S, u) \left(
\egamma + \log2 - i\eta_1 (u+i/2) \right. \\ \left. - H_1 (S) + i \pi \tanh (\pi u)
\right) + \sum_{k=1}^S \frac{1+(-1)^k}{k} Q (S-k, u)
\Big\} \\
- \alpha \sum_{k=1}^S \frac{1+(-1)^k}{k} \left(
B_1 (S) {+} B_1 (S-k)
\right) Q (S-k, u) \\ + \phi_{1,0}^{\mathrm{per}} Q (S, u)
+ \phi_{1,1}^{\mathrm{per}} \P_1 (u+i/2) Q (S, u)\,.
\end{multline}

The solution of second Baxter equation at NLO goes along the same lines, but the expressions become more cumbersome. So, here we will present only the expression for $q_2^{(1)}(u)$ without derivation. The latter including homogeneous piece with at maximum first order poles in $u$ as required by QSC analyticity constraints is given by 
\begin{multline}
\frac{q_2^{(1)}(u)}{\alpha} = 
A_3^{(1)} \left(\frac{\sigma  \left(3 S^2+3 S+1\right) Q_S}{4 (2 S+1)}-\frac{3 \sigma  (S-1) S Q_{S-2}}{8 (2 S-1)}-\frac{3
	\sigma  (S+1) (S+2) Q_{S+2}}{8 (2 S+3)}\right) \\ + 2 (2 S+3) B_1(S) \left\{\left(\frac{1}{2 \left\lfloor \frac{S+1}{2}\right\rfloor -1}+\sigma
\right) Q_{S-2} + \frac{(1-\sigma  (2 S-1)) Q_S}{2 S+1}\right\} -4 B_1(S) G_3(S) \\ -\frac{k_1^{(0)} \sigma  (2 S+3) B_1(S) Q_S}{(2
	S+1)^2} +\left(\frac{\sigma  (2 S-1) (2 S+3) \left(3 S^2+3 S+1\right) B_1(S)}{3 (2
	S+1)}-\frac{k_1^{(0)} \sigma }{2 (2 S+1)}\right) \\ \times \left\{2 B_1(S) G_6(S)+G_4(S)-\alpha^{-1}\Phi_1^{\mathrm{per}} (u) Q_S \right\} +\frac{1}{4} i k_2^{(1)} \sigma  \left(Q_{S+1}-\delta
_{S\neq 0} Q_{S-1}\right) \\ -\frac{1}{2} \sigma  (S-1) S (2 S+3) B_1(S) \left\{(2 B_1(S)
G_6(S-2)+G_4(S-2)-\alpha^{-1}\Phi_1^{\mathrm{per}} (u) Q_{S-2}\right\} \\ -\frac{1}{2} \sigma  (S+1) (S+2) (2 S-1) B_1(S) \left\{ 2 B_1(S) G_6(S+2)+G_4(S+2)-\alpha^{-1}\Phi_1^{\mathrm{per}} (u) Q_{S+2}\right\} \\ +\sigma  (2 S-1) (2 S+3) B_1(S){}^2 \Bigg\{\frac{\left(6 S^3-3 S^2-11 S-3\right) Q_S}{3 (2 S+1)^2}-\frac{\left(2 S^3-3 S^2-23
S-22\right) Q_{S+2}}{2 (2 S+3)^2} \\ -\frac{(S-1) S (2 S+3) Q_{S-2}}{2 (2 S-1)^2}\Bigg\} + \frac{k_1^{(1)} \sigma  Q_S}{2 (2 S+1)} + \alpha^{-1}\phi_{2,0}^{\mathrm{per}} \mathcal{Z}_S + \alpha^{-1}\phi_{2,0}^{\mathrm{anti}} \P_{-1} \left(u+\tfrac i2\right) Q_S\, , \label{q21}
\end{multline}
where $Q_S\equiv Q_S (u)\equiv Q (S, u)$, $\mathcal{Z}_S\equiv \mathcal{Z} (S, u)$ and the following functions were introduced:
\begin{align}
G_3 (S)\equiv G_3 (S,u) =& \sum_{k=S+1}^{2 S+1} \frac{(1-(-1)^k) Q_{k-S-1}(u)}{k}\, , \\
G_4 (S)\equiv G_4 (S,u) =& -\sum_{k=1}^S \frac{(1+(-1)^k) Q_{S-k} (u)}{k}\left(
B_1 (S) - B_1 (S-k)
\right)\, , \\
G_6 (S)\equiv G_6 (S,u) =&  -\sum_{k=1}^S \frac{(1+(-1)^k) Q_{S-k} (u)}{k} \\ & - 2 Q_S (u) \Big\{\gamma_E + \log 2
- H_1 (S) - i\eta_1 (u+i/2) + i\pi\tanh (\pi u)
\Big\}\, .
\end{align}

\section{Anomalous dimensions}\label{AnomalousDimensions}
\noindent
The expressions for the anomalous dimensions could be easily obtained from $A_3^{(0,1)}$ with the help of (\ref{Pasympt}). This way, up to four loops we got the following results for anomalous dimensions of twist 1 operators
\begin{equation}
\gamma (S) = \gamma^{(0)} (S) h^2 + \gamma^{(1)} (S) h^4 + \ldots
\end{equation}
where
\begin{equation}
\gamma^{(0)}(S) = 4 \left(
H_1 - H_{-1}\right)\,,
\end{equation}
\begin{multline}
\gamma^{(1)}(S) = 16 \Big\{ 3 \barH_{-2,-1} - 2 \barH_{-2, i} - \barH_{-2,1} - \barH_{-1,-2} + 2 \barH_{-1, 2 i} - \barH_{-1,2} - 6 \barH_{i,-2} \\
+ 12 \barH_{i, 2 i} - 6 \barH_{i, 2} - 6 \barH_{2 i, -1} + 4 \barH_{2 i, i} + 2 \barH_{2 i, 1} - \barH_{1,-2} + 2 \barH_{1, 2 i} - \barH_{1,2} + 3\barH_{2,-1} \\ - 2 \barH_{2, i} - \barH_{2,1} + 2\barH_{-1, i, -1} - 2 \barH_{-1, i, 1} + 8\barH_{i, -1, -1} - 12\barH_{i, -1, i} + 4 \barH_{i, -1, 1} - 16\barH_{i, i, -1} \\
+ 16 \barH_{i, i, i} + 4 \barH_{i, 1, -1} - 4 \barH_{i, 1, i} + 2 \barH_{1,i,-1} - 2 \barH_{1,i,1} \Big\} + 8 \left(
H_{-1} - H_1
\right)\zeta_2\,,\label{eq:gammaNLO} \\
\end{multline}
and we introduced new sums 
\begin{equation}\label{eq:sums}
H_{a,b,\ldots}(S) = \sum_{k=1}^S \frac{\Re [(a/|a|)^k]}{k^{|a|}} H_{b,\ldots} (k)\,, \quad H_{a,\ldots} = H_{a,\ldots} (S)\,, \quad \barH_{a,\ldots} = H_{a,\ldots} (2S)\,,
\end{equation}
so that for real indexes these sums reduce to ordinary harmonic sums. Note that the conventional harmonic sums of argument $S$ can be expressed via our extended basis \eqref{eq:sums}. In particular, we have $\gamma^{(0)}(S) = 4 \left(
H_1 - H_{-1}\right) = 4\left(\barH_1 + \barH_{-1}-2\barH_i\right)$.

Imaginary indexes correspond to the generalization of the harmonic sums with the fourth root of unity factor $(\exp(i\pi/2))^n$. We would like also to point out the appearance of argument $2S$ and note, that these new sums could not be further reduced  to cyclotomic or other generalized harmonic sums of \cite{Ablinger3,Ablinger4}. From this expression we see, that the maximal transcendentality principle\footnote{Similar considerations for the evaluation of Feynman diagrams first appeared in \cite{KotikovVeretin}. See also \cite{Tolya1,Tolya2}.} \cite{N4SYM2loop4,N4SYM3loop} also holds for anomalous dimensions of ABJM theory with the account for finite size corrections. Next, the obtained expression respects shift symmetry \cite{Beccaria1,ABJMQSC12loops}, that is maximum transcendentality parts ($\zeta_2$ pieces in our case) of anomalous dimensions of operators with $(L,S) = (1,2n)$ and $(L,S) = (1,2n-1)$, $n\in \mathbb{N}^+$ are the same. It is also interesting to inspect the large spin $S$ limit of our expressions. In this limit we get 
\begin{align}
\gamma^{(0)}(S) &\sim 4\log (S) + 4 \left(
\log2 + \gamma_E
\right)  + \OO (\frac{1}{S})\,, \\
\gamma^{(1)}(S) &\sim -8\zeta_2\log (S) - 8\zeta_2 \left(
\log2 + \gamma_E
\right) - 12\zeta_3 + \OO (\frac{\log (S)}{S})\,,
\end{align}
which is in agreement with \cite{Beccaria1,Beccaria2}, where it was shown that wrapping correction scales as $\log S / S$. 

Let us now make some comments on the size of the basis of generalized harmonic sums \eqref{eq:sums} at NNLO and higher.  At weight $w$ the number of all such sums is given by $4\cdot 5^{w-1}$ ($3\cdot 4^{w-1}$ if we take into account the real part operator in \eqref{eq:sums}). However, similar to S-sums \cite{S-sums1,S-sums2} these sums should obey additional relations similar to those following from quasi-shuffle algebra together with  differentiation and generalized argument relations for S-sums. Indeed at NLO with weight 3 we expect in total 100 (48) such sums. However, only 29 of them appear in our final answer (\ref{eq:gammaNLO}). Similarly, at NNLO with $w=5$ in general we should have 2500 (768) sums which are expected to further reduce to several hundreds. Also, in the case of ABJM model compared to $\mathcal{N}=4$ SYM the analyticity properties of anomalous dimensions are fully unstudied at the moment and we can not use for example such notion as Gribov-Lipatov reciprocity \cite{GL-reciprocity1,GL-reciprocity2} to further reduce sums basis, see for example \cite{N4SYM7loop}.

Finally, it is instructive to compare our result with previously known expression \cite{Beccaria1,Beccaria2}:
\begin{multline}
\gamma^{L=1, S} = 4 h^2 \left(
H_1 - H_{-1}
\right) - 16 h^4 \Big( H_{-3} - H_3 + H_{-2,-1} - H_{-2,1} + H_{-1,-2} - H_{-1,2} \\ - H_{1,-2} + H_{1,2} - H_{2,-1} + H_{2,1} + H_{-1,-1,-1} - H_{-1,-1,1} - H_{1,-1,-1} + H_{1,-1,1} \Big) \\
+ 4 h^4 \left(
H_1 - H_{-1}
\right) \mathcal{W} (1,S) + \OO (h^6)\, ,
\end{multline}
where
\begin{multline}
\mathcal{W} (L, S) = -\frac{i}{2}\sum_{M=1}^{\infty} (-1)^M \operatornamewithlimits{Res}_{q=iM} \frac{4^L Q\left(\frac{q-i (M-1)}{2}\right)}{(q^2+M^2)^L Q\left(
\frac{q-i (M+1)}{2}	
\right) Q\left(
\frac{q+ i (M-1)}{2}
\right)  Q\left(
\frac{q+i (M+1)}{2}
\right)} \\
\times \sum_{j=0}^{M-1}\left[
\frac{Q\left(\frac{q-i (M-1)+2 i j}{2}\right)}{Q \left(
\frac{q- i (M-1)}{2}	
\right)}
\right]^2 \left(
\frac{1}{2 j - i q - M} - \frac{1}{2 (j+1)-i q - M}
\right)
\end{multline}
and $Q (u) = \, _2 F_1 (-S, i u + 1/2; 1; 2)$. This expression is still too complex as it involves "complex" hypergeometric functions and takes more time to get expressions for anomalous dimensions compared to our representation in terms of generalized harmonic sums. Moreover the representation in terms of sums makes the study of analytic properties of anomalous dimensions much more simple.

\section{Conclusion}\label{Conclusion}

In this paper we have shown how Mellin space technique could be used to solve multiloop Baxter equations arising from $\mathcal{N}=4$ SYM or ABJM quantum spectral curves. As a particular example we have considered anomalous dimensions of twist 1 operators in ABJM theory up to four loop order. The result could be expressed in terms of harmonic sums with imaginary indexes, so that the maximum transcendentality principle holds. The presented Mellin space technique for the solution of multiloop Baxter equations arising within quantum spectral curve method could be extended to higher loops at twist 1. Moreover, it could be also applied to the twist 2 operators. In the latter case we will have an inhomogeneous second order differential equation in Mellin space, which could be further solved using Abel's reduction of order to find second homogeneous solution\footnote{The first homogeneous solution is known for arbitrary spin values.} and variation of constants to determine particular solution of inhomogeneous equation. However, the main technical difficulty within the presented approach is related to finding Mellin and inverse Mellin transforms of arising functions. Our preliminary results on the solution of multiloop Baxter equations directly in spectral parameter $u$-space for arbitrary spin values show that such a direct approach could be much more effective both for having results at higher loops and twists. So, next we are planning to concentrate on developing corresponding $u$-space techniques. 

Finally, we would like to note that the presented techniques could be also applied for solving twisted $\mathcal{N}=4$ and ABJM quantum spectral curves, where $\bP$ function will have twisted non-polynomial asymptotic at large spectral parameter values. The latter models are interesting in the connection with the recent progress with the so called fishnet theories  \cite{fishnet1,Isaev,fishnet2,fishnet3,fishnet4,fishnet5,fishnet6,fishnet7}.

\section*{Acknowledgements}

This work was supported by RFBR grants \# 17-02-00872, \# 16-02-00943 and contract \# 02.A03.21.0003 from 27.08.2013 with Russian Ministry of Science and Education. The work of R. Lee was supported by the grant of the “Basis” foundation for theoretical physics.

\appendix

\section{Mellin transformation}\label{MellinTransformation}

We define Mellin transformation and its inverse  similar to \cite{Kotikov2}, that is 
\begin{align}
f\left(u\right) & =\mathcal{M}[\tilde{f}]\left(u\right) =\frac{1}{K_{1}\left(u\right)}\intop_{0}^{1}dz\,z^{iu-\frac{1}{2}}\bar{z}^{-iu-\frac{1}{2}}\tilde{f}\left(z\right)\,,\\
\tilde{f}\left(z\right) &=\mathcal{M}^{-1}[f]\left(z\right) =\frac{1}{2\pi}\intop_{-\infty}^{\infty}du\,z^{-iu-\frac{1}{2}}\bar{z}^{iu-\frac{1}{2}}K_{1}\left(u\right)f\left(u\right)\,,
\end{align} 
where
\begin{equation}
K_{1}\left(u\right)=\Gamma\left(\tfrac{1}{2}+iu\right)\,\Gamma\left(\tfrac{1}{2}-iu\right)=\frac{\pi}{\cosh\left(\pi u\right)}\,,
\end{equation}
and $\bar{z}=1-z$.
Next, it is easy to see, that the introduced transformations have the following properties
\begin{equation}
\widetilde{uf\left(u\right)}\left(z\right) =\hat{u}\tilde{f}\left(z\right)=\frac{i}{2}\left(\bar{z}-z+2z\bar{z}\partial_{z}\right)\tilde{f}\left(z\right)\,,
\end{equation}
\begin{equation}\label{eq:shiftup}
\widetilde{f\left(u+i\right)}\left(z\right) =-\frac{\bar{z}}{z}\tilde{f}\left(z\right){+\frac{1}{z}f\left(i/2\right)}=-\frac{\bar{z}}{z}\tilde{f}\left(z\right){+\frac{1}{z}\tilde{f}(0)},
\end{equation}
\begin{equation}\label{eq:shiftdown}
\widetilde{f\left(u-i\right)}\left(z\right) =-\frac{z}{\bar{z}}\tilde{f}\left(z\right){+\frac{1}{\bar{z}}f\left(-i/2\right)}
=-\frac{z}{\bar{z}}\tilde{f}\left(z\right){+\frac{1}{\bar{z}}\tilde{f}\left(1\right)}\,. 
\end{equation}
Eqs. \eqref{eq:shiftup} and \eqref{eq:shiftdown} are valid when $f\left(u\right)$ does not have singularities in the strip $0<\Im u<1$ and $-1<\Im u<0$, respectively.
It is also convenient to introduce the following shorthand notations widely used in the main body of the paper
\begin{align}\label{eq:hat_u}
\hat{u} & =\frac{i}{2}\left(\bar{z}-z+2z\bar{z}\partial_{z}\right)=\sqrt{z\bar{z}}i\partial_{z}\sqrt{z\bar{z}}\,,\\
f^{\left[n\right]}\left(u\right) & =f\left(u+in/2\right)\,.
\end{align}
The Mellin transform and its inverse convert polynomials into polynomials. Moreover, polynomials of degree $k$ are converted into polynomials with the same degree. That is
\begin{equation}
\frac{1}{K_{1}\left(u\right)}\intop_{0}^{1}dz\,z^{iu-\frac{1}{2}}\bar{z}^{-iu-\frac{1}{2}}z^{k}=\frac{\Gamma\left(\frac{1}{2}+iu+k\right)\,\Gamma\left(\frac{1}{2}-iu\right)}{K_{1}\left(u\right)k!}=\frac{\left(\frac{1}{2}+iu\right)_{k}}{k!}
\end{equation}
and 
\begin{equation}
\frac{1}{2\pi}\intop_{-\infty}^{\infty}du\,z^{-iu-\frac{1}{2}}\bar{z}^{iu-\frac{1}{2}}K_{1}\left(u\right)u^{k}  =\hat{u}^k1=\left[\frac{i}{2}\left(\bar{z}-z+2z\bar{z}\partial_{z}\right)\right]^{k}1\,.
\end{equation}

\section{Generating function and properties of Baxter polynomials}\label{BaxterDerivatives}

Let us define the generating function for Baxter polynomials as
\begin{equation}
	W(x,u)=\sum_{S=0}^\infty Q(S,u)x^S\,.
\end{equation}
In order to obtain the differential equation for $W(x,u)$, we use the relation
\begin{multline}
-a^{2}\,_{2}F_{1}(a+1,b;a+b+1;-1)+(a+b-1)(a+b)\,_{2}F_{1}(a-1,b;a+b-1;-1)\\
-(2b-1)(a+b)\,_{2}F_{1}(a,b;a+b;-1)=0\,.
\end{multline}
Putting $a=-S,\,b=\frac{1}{2}+iu$ we obtain the relation between the Baxter polynomials\footnote{There is another way to find this relation by noting the formal symmetry of the definition \eqref{eq:BaxterPolynomial} with respect to the substitution $-S\leftrightarrow 1/2+iu$ and applying this substitution to the homogeneous part of Baxter equation \eqref{Baxter1a}.}
\begin{equation}
SQ\left(S-1,u\right)-\left(S+1\right)Q\left(S+1,u\right)-2iuQ\left(S,u\right)=0\,.
\end{equation}

Multiplying the above recurrence relation with $x^{S}$  and summing over $S$, we obtain
\begin{equation}
x\partial_{x}xW\left(x,u\right)-\partial_{x}W\left(x,u\right)-2iuW\left(x,u\right)=0\,.
\end{equation}

Solving this differential equation with the boundary condition $W\left(0,u\right)=Q\left(0,u\right)=1$,
we have
\begin{equation}\label{eq:W}
W\left(x,u\right)=(1-x)^{-\frac{1}{2}+iu}(1+x)^{-\frac{1}{2}-iu}=\frac{e^{-2i(u-i/2)\, \mathrm{arctanh}\,x}}{1-x}\,,
\end{equation}
We will use this generating function $W\left(x,u\right)$ for two purposes. First, let us prove that 
\begin{equation}\label{eq:Z}
\Z (S,u) = i\sigma\sum_{k=0}^{\left\lfloor \frac{S-1}{2}\right\rfloor }\frac{1}{S-k}Q\left(S-1-2k,u\right)+\sigma\eta_{-1}(u+{i}/{2})Q\left(S,u\right)
\end{equation}
is a solution of the homogeneous part of the second Baxter equation \eqref{Baxter2a}, i.e., that
\begin{equation}
	(u+i/2)\mathcal{Z}(S,u+i)+i(2S+1)\mathcal{Z}(S,u)-(u-i/2)\mathcal{Z}(S,u-i)=0\,.
\end{equation}
Substituting \eqref{eq:Z} in the left-hand side, we obtain
\begin{multline}\label{eq:Zeq} 
	(u+i/2)\mathcal{Z}(u+i)+i(2S+1)\mathcal{Z}(u)-(u-i/2)\mathcal{Z}(u-i)=Q\left(S,u+i\right)-Q\left(S,u-i\right)\\
	+2i\sum_{k=0}^{S-1}\frac{1-\left(-1\right)^{k+S}}{2k+1}\left[\left(u+i/2\right)Q\left(k,u+i\right)-\left(u-i/2\right)Q\left(k,u-i\right)\right]
\end{multline}
We have used here the first Baxter equation to express $Q(S,u)$ and  $Q(S-1-2k,u)$ via $Q(S,u\pm i)$ and  $Q(S-1-2k,u\pm i)$, respectively. We also used the identity $\eta_{-1}(u)+\eta_{-1}(u+i)=u^{-1}$. Now, to prove that the right-hand side of Eq. \eqref{eq:Zeq} is zero, we sum it over $S$ with the weight $x^S$. Then we have  
\begin{multline}
\sum_{S=0}^{\infty}x^S[\mbox{r.-h.s. of \eqref{eq:Zeq}}]=	W\left(x,u+i\right)-W\left(x,u-i\right)\\
+2i\sum_{k=0}^{\infty}\frac{1}{2k+1}\left[\left(u+i/2\right)Q\left(k,u+i\right)-\left(u-i/2\right)Q\left(k,u-i\right)\right]\sum_{S=k+1}^{\infty}x^{S}\left(1-\left(-1\right)^{k+S}\right)\\
=W\left(x,u+i\right)-W\left(x,u-i\right)+\frac{2i\sqrt{x}}{1-x^{2}}\intop_0^x\frac{dy}{\sqrt{y}}\left[\left(u+\tfrac{i}2\right)W\left(y,u+i\right)-\left(u-\tfrac{i}2\right)W\left(y,u-i\right)\right].
\end{multline}
Then, using the explicit form of $W(x,u)$, Eq. \eqref{eq:W}, we can check that the last expression is zero.

Now, let us derive a closed expression for the derivatives of Baxter polynomials at $u=i/2$. From \eqref{eq:W} we have
\begin{equation}
\partial_{u}^{n}W\left(x,u\right)|_{u=i/2}=\frac{\left(-2i\,\mathrm{arctanh}\,x\right)^{n}}{1-x}\,.
\end{equation}
The coefficient in front of $x^{S}$ is the $n$-th derivative of $Q\left(S,u\right)$
at $u=i/2$. Using $\frac{1}{1-x}=\sum x^{n}$ we obtain
\begin{equation}
Q^{\left(n\right)}\left(S,i/2\right)=\left(-2i\right)^{n}\left.\left\langle \left(\mathrm{arctanh}\,x\right)^{n}\right\rangle _{S}\right|_{x=1},
\end{equation}
where
\begin{equation}
\left\langle f\left(x\right)\right\rangle _{S}=\sum_{k=0}^{S}\frac{f^{\left(k\right)}\left(0\right)}{k!}x^{k}
\end{equation}
denotes the Taylor expansion of the function $f(x)$ interrupted at $S$-th term. Using the Taylor expansion of $\mathrm{arctanh}\,x$, we have
\begin{equation}
Q^{\left(n\right)}\left(S,i/2\right)=\left(-i\right)^{n}\sum_{\substack{
	i_{1},\ldots ,i_{n}\\i_{1}+\ldots+i_{n}\leqslant\frac{S-n}{2}}}\prod_{k=1}^{n}\frac{1}{i_{k}+\frac{1}{2}}\,.
\end{equation}
After a little algebra we obtain
\begin{equation}
i^{n}Q^{\left(n\right)}\left(S,i/2\right)=\sum_{N=0}^{S-n}\frac{n\left[1+(-1)^N\right]}{N+n}i^{n-1}Q^{\left(n-1\right)}\left(N+n-1,i/2\right)\,.
\end{equation}
Since $Q^{\left(n\right)}\left(S,i/2\right)=0$ for $S<n$, we can replace the lower limit with $-n+1$ and then shift $N\to N-n$. 
Then we have
\begin{align}
\frac{i^{n}}{n!}Q^{\left(n\right)}\left(S,i/2\right)=\sum_{N=1}^{S}\frac{1+(-1)^{N+n}}{N}\frac{i^{n-1}}{(n-1)!}Q^{\left(n-1\right)}\left(N+n-1,i/2\right)\,.
\end{align}
Finally, we obtain
\begin{align}
Q^{\left(n\right)}\left(S,i/2\right)=(-i)^{n}n! B_n(S),
\end{align}
where $B_0(S)=1$, $B_1(S)=H_1(S)-H_{-1}(S)$, and $B_{n>1}$ is defined recursively by the symbolic formula
\begin{equation}
B_{n}=\left(O_{1}+\left(-1\right)^{n}O_{-1}\right)B_{n-1}\,,
\end{equation}
where $O_{\pm1}$ is a linear operator prepending index $\pm 1$ to harmonic sums, i.e. $O_{\pm1}H_{\boldsymbol{a}}\left(S\right)=H_{\pm1,\boldsymbol{a}}\left(S\right)$.
In particular, we have
\begin{align}
B_{2}=&\left(O_{1}+O_{-1}\right)B_{1}=H_{1,1}+H_{-1,1}-H_{1,-1}-H_{-1,-1}\,,\\
B_{3}=&\left(O_{1}-O_{-1}\right)B_{2}=H_{1,1,1}+H_{1,-1,1}-H_{1,1,-1}-H_{1,-1,-1}\\
&-H_{-1,1,1}-H_{-1,-1,1}+H_{-1,1,-1}+H_{-1,-1,-1}\,,
\end{align}
where we omitted the argument $S$ for brevity.

\section{Mellin transformations of arising functions}\label{MellinForFunctions}

Let us present inverse Mellin images of the functions which arise in $z$-space.
 
\begin{multline}\label{eq:wlog}
\mathcal{M}\left[w^S\log|w|\right] =Q\left(S,u\right)\left[\egamma+\log2-H_1\left(S\right)-i\eta_{1}\left(u+\frac{i}{2}\right)-\sinh(\pi u)\eta_{-1}\left(u+\frac{i}{2}\right)\right]\\
-i\sinh(\pi u)\sum_{k=1}^{S}\frac{1-(-1)^k}{2S-k+1}Q(S-k,u)+\sum_{k=1}^S\frac{1+(-1)^k}{k} Q (S-k, u)\,, 
\end{multline}
where $w = \bar z - z$. Note that the combination $f(u)=-i\eta_{1}\left(u+\frac{i}{2}\right)-\sinh(\pi u)\eta_{-1}\left(u+\frac{i}{2}\right)$ appearing in the above formula is an even function, $f(u)=f(-u)$.


Similarly we obtained
\begin{multline}
\mathcal{M}\left[w^S\log (1-w)\right] = Q (S, u)
\left[\egamma + \log2 -i\eta_1 (-u+\frac{i}{2}) - H_1 (S)\right]
+ \sum_{k=1}^S\frac{(-1)^k}{k} Q(S-k, u)
\end{multline}
and 
\begin{multline}
\mathcal{M}\left[w^S\log (1+w)\right] = Q(S,u)\left[
\egamma + \log2 - i\eta_1 (u+\frac{i}{2}) - H_1 (S)
\right] + \sum_{k=1}^S\frac{1}{k} Q (S-k, u)\,.
\end{multline}

During the calculation we also needed the expression for inverse Mellin transform of $\mathcal{P}_1(u+\frac{i}{2}) Q(S,u) = \pi \tanh(\pi u) Q(S,u)$:
\begin{equation}
	\mathcal{M}^{-1}\left[\pi \tanh(\pi u) Q(S,u)\right]= i\ln\left((1+w)/(1-w)\right)w^{S}-i\sum_{k=1}^{S}\frac{1-(-1)^k}{k}w^{S-k}\,.
\end{equation}

\bibliographystyle{hieeetr}
\bibliography{litr}

\begin{thebibliography}{100}

\bibitem{tHooftDuality}
G.~'t~Hooft, ``{A Planar Diagram Theory for Strong Interactions},'' {\em Nucl.
  Phys.}, vol.~B72, p.~461, 1974.

\bibitem{MaldacenaAdSCFT}
J.~M. Maldacena, ``{The Large N limit of superconformal field theories and
  supergravity},'' {\em Int. J. Theor. Phys.}, vol.~38, pp.~1113--1133, 1999,
  hep-th/9711200.
\newblock [Adv. Theor. Math. Phys.2,231(1998)].

\bibitem{GKP}
S.~S. Gubser, I.~R. Klebanov, and A.~M. Polyakov, ``{Gauge theory correlators
  from noncritical string theory},'' {\em Phys. Lett.}, vol.~B428,
  pp.~105--114, 1998, hep-th/9802109.

\bibitem{WittenAdSHolography}
E.~Witten, ``{Anti-de Sitter space and holography},'' {\em Adv. Theor. Math.
  Phys.}, vol.~2, pp.~253--291, 1998, hep-th/9802150.

\bibitem{IntegrabilityReview}
N.~Beisert {\em et~al.}, ``{Review of AdS/CFT Integrability: An Overview},''
  {\em Lett. Math. Phys.}, vol.~99, pp.~3--32, 2012, arXiv:1012.3982.

\bibitem{IntegrabilityPrimer}
D.~Bombardelli, A.~Cagnazzo, R.~Frassek, F.~Levkovich-Maslyuk, F.~Loebbert,
  S.~Negro, I.~M. Szécsényi, A.~Sfondrini, S.~J. van Tongeren, and
  A.~Torrielli, ``{An integrability primer for the gauge-gravity
  correspondence: An introduction},'' {\em J. Phys.}, vol.~A49, no.~32,
  p.~320301, 2016, arXiv:1606.02945.

\bibitem{IntegrabilityDeformations}
S.~J. van Tongeren, ``{Integrability of the ${\rm Ad}{{{\rm S}}_{5}}\times
  {{{\rm S}}^{5}}$ superstring and its deformations},'' {\em J. Phys.},
  vol.~A47, p.~433001, 2014, arXiv:1310.4854.

\bibitem{IntegrabilityDefects}
M.~de~Leeuw, A.~C. Ipsen, C.~Kristjansen, and M.~Wilhelm, ``{Introduction to
  Integrability and One-point Functions in $\mathcal{N}=4$ SYM and its Defect
  Cousin},'' in {\em {Les Houches Summer School: Integrability: From
  Statistical Systems to Gauge Theory Les Houches, France, June 6-July 1,
  2016}}, 2017, arXiv:1708.02525.

\bibitem{IntroductionQSC}
N.~Gromov, ``{Introduction to the Spectrum of $N=4$ SYM and the Quantum
  Spectral Curve},'' 2017, arXiv:1708.03648.

\bibitem{IntegrabilityStructureConstants}
S.~Komatsu, ``{Lectures on Three-point Functions in N=4 Supersymmetric
  Yang-Mills Theory},'' 2017, arXiv:1710.03853.

\bibitem{ABJM1}
O.~Aharony, O.~Bergman, D.~L. Jafferis, and J.~Maldacena, ``{N=6 superconformal
  Chern-Simons-matter theories, M2-branes and their gravity duals},'' {\em
  JHEP}, vol.~10, p.~091, 2008, arXiv:0806.1218.

\bibitem{StaudacherSMatrix}
M.~Staudacher, ``{The Factorized S-matrix of CFT/AdS},'' {\em JHEP}, vol.~05,
  p.~054, 2005, hep-th/0412188.

\bibitem{BetheAnsatzQuantumStrings}
G.~Arutyunov, S.~Frolov, and M.~Staudacher, ``{Bethe ansatz for quantum
  strings},'' {\em JHEP}, vol.~10, p.~016, 2004, hep-th/0406256.

\bibitem{BeisertDynamicSmatrix}
N.~Beisert, ``{The $SU(2|2)$ dynamic S-matrix},'' {\em Adv. Theor. Math.
  Phys.}, vol.~12, pp.~945--979, 2008, hep-th/0511082.

\bibitem{BeisertAnalyticBetheAnsatz}
N.~Beisert, ``{The Analytic Bethe Ansatz for a Chain with Centrally Extended
  $su(2|2)$ Symmetry},'' {\em J. Stat. Mech.}, vol.~0701, p.~P01017, 2007,
  nlin/0610017.

\bibitem{TranscedentalityCrossing}
N.~Beisert, B.~Eden, and M.~Staudacher, ``{Transcendentality and Crossing},''
  {\em J. Stat. Mech.}, vol.~0701, p.~P01021, 2007, hep-th/0610251.

\bibitem{JanikWorldsheetSmatrix}
R.~A. Janik, ``{The AdS(5) x S**5 superstring worldsheet S-matrix and crossing
  symmetry},'' {\em Phys. Rev.}, vol.~D73, p.~086006, 2006, hep-th/0603038.

\bibitem{ArutyunovFrolovSmatrix}
G.~Arutyunov and S.~Frolov, ``{On AdS(5) x S**5 String S-matrix},'' {\em Phys.
  Lett.}, vol.~B639, pp.~378--382, 2006, hep-th/0604043.

\bibitem{ZamolodchikovFaddevAlgebraAdS}
G.~Arutyunov, S.~Frolov, and M.~Zamaklar, ``{The Zamolodchikov-Faddeev algebra
  for AdS(5) x S**5 superstring},'' {\em JHEP}, vol.~04, p.~002, 2007,
  hep-th/0612229.

\bibitem{N6Smatrix}
C.~Ahn and R.~I. Nepomechie, ``{N=6 super Chern-Simons theory S-matrix and
  all-loop Bethe ansatz equations},'' {\em JHEP}, vol.~09, p.~010, 2008,
  arXiv:0807.1924.

\bibitem{MinahanZarembo}
J.~A. Minahan and K.~Zarembo, ``{The Bethe ansatz for N=4 superYang-Mills},''
  {\em JHEP}, vol.~03, p.~013, 2003, hep-th/0212208.

\bibitem{N4SuperSpinChain}
N.~Beisert and M.~Staudacher, ``{The N=4 SYM integrable super spin chain},''
  {\em Nucl. Phys.}, vol.~B670, pp.~439--463, 2003, hep-th/0307042.

\bibitem{LongRangeBetheAnsatz}
N.~Beisert and M.~Staudacher, ``{Long-range $psu(2,2|4)$ Bethe Ansatze for
  gauge theory and strings},'' {\em Nucl. Phys.}, vol.~B727, pp.~1--62, 2005,
  hep-th/0504190.

\bibitem{MinahanZaremboChernSimons}
J.~A. Minahan and K.~Zarembo, ``{The Bethe ansatz for superconformal
  Chern-Simons},'' {\em JHEP}, vol.~09, p.~040, 2008, arXiv:0806.3951.

\bibitem{SpinChainsN6ChernSimons}
D.~Gaiotto, S.~Giombi, and X.~Yin, ``{Spin Chains in N=6 Superconformal
  Chern-Simons-Matter Theory},'' {\em JHEP}, vol.~04, p.~066, 2009,
  arXiv:0806.4589.

\bibitem{AllloopAdS4}
N.~Gromov and P.~Vieira, ``{The all loop AdS4/CFT3 Bethe ansatz},'' {\em JHEP},
  vol.~01, p.~016, 2009, arXiv:0807.0777.

\bibitem{TBAN4}
N.~Gromov, V.~Kazakov, and P.~Vieira, ``{Exact Spectrum of Anomalous Dimensions
  of Planar N=4 Supersymmetric Yang-Mills Theory},'' {\em Phys. Rev. Lett.},
  vol.~103, p.~131601, 2009, arXiv:0901.3753.

\bibitem{TBAN4proposal}
D.~Bombardelli, D.~Fioravanti, and R.~Tateo, ``{Thermodynamic Bethe Ansatz for
  planar AdS/CFT: A Proposal},'' {\em J. Phys.}, vol.~A42, p.~375401, 2009,
  arXiv:0902.3930.

\bibitem{TBAexcitedstates}
N.~Gromov, V.~Kazakov, A.~Kozak, and P.~Vieira, ``{Exact Spectrum of Anomalous
  Dimensions of Planar N = 4 Supersymmetric Yang-Mills Theory: TBA and excited
  states},'' {\em Lett. Math. Phys.}, vol.~91, pp.~265--287, 2010,
  arXiv:0902.4458.

\bibitem{TBAMirrorModel}
G.~Arutyunov and S.~Frolov, ``{Thermodynamic Bethe Ansatz for the AdS(5) x S(5)
  Mirror Model},'' {\em JHEP}, vol.~05, p.~068, 2009, arXiv:0903.0141.

\bibitem{YsystemAdS5}
A.~Cavaglia, D.~Fioravanti, and R.~Tateo, ``{Extended Y-system for the
  $AdS_5/CFT_4$ correspondence},'' {\em Nucl. Phys.}, vol.~B843, pp.~302--343,
  2011, arXiv:1005.3016.

\bibitem{TBAfromYsystem}
J.~Balog and A.~Hegedus, ``{$AdS_5\times S^5$ mirror TBA equations from
  Y-system and discontinuity relations},'' {\em JHEP}, vol.~08, p.~095, 2011,
  arXiv:1104.4054.

\bibitem{WronskianSolution}
N.~Gromov, V.~Kazakov, S.~Leurent, and Z.~Tsuboi, ``{Wronskian Solution for
  AdS/CFT Y-system},'' {\em JHEP}, vol.~01, p.~155, 2011, arXiv:1010.2720.

\bibitem{SolvingYsystem}
N.~Gromov, V.~Kazakov, S.~Leurent, and D.~Volin, ``{Solving the AdS/CFT
  Y-system},'' {\em JHEP}, vol.~07, p.~023, 2012, arXiv:1110.0562.

\bibitem{TBAYsystemAdS4}
D.~Bombardelli, D.~Fioravanti, and R.~Tateo, ``{TBA and Y-system for planar
  AdS(4)/CFT(3)},'' {\em Nucl. Phys.}, vol.~B834, pp.~543--561, 2010,
  arXiv:0912.4715.

\bibitem{GromovYsystemAdS4}
N.~Gromov and F.~Levkovich-Maslyuk, ``{Y-system, TBA and Quasi-Classical
  strings in AdS(4) x CP3},'' {\em JHEP}, vol.~06, p.~088, 2010,
  arXiv:0912.4911.

\bibitem{DiscontinuityRelationsAdS4}
A.~Cavaglia, D.~Fioravanti, and R.~Tateo, ``{Discontinuity relations for the
  $AdS_4/CFT_3$ correspondence},'' {\em Nucl. Phys.}, vol.~B877, pp.~852--884,
  2013, arXiv:1307.7587.

\bibitem{potentialTBA}
D.~Correa, J.~Maldacena, and A.~Sever, ``{The quark anti-quark potential and
  the cusp anomalous dimension from a TBA equation},'' {\em JHEP}, vol.~08,
  p.~134, 2012, arXiv:1203.1913.

\bibitem{IntegrableWilsonLoops}
N.~Drukker, ``{Integrable Wilson loops},'' {\em JHEP}, vol.~10, p.~135, 2013,
  arXiv:1203.1617.

\bibitem{cuspQSC}
N.~Gromov and F.~Levkovich-Maslyuk, ``{Quantum Spectral Curve for a cusped
  Wilson line in $ \mathcal{N}=4 $ SYM},'' {\em JHEP}, vol.~04, p.~134, 2016,
  arXiv:1510.02098.

\bibitem{potentialQSC}
N.~Gromov and F.~Levkovich-Maslyuk, ``{Quark-anti-quark potential in $
  \mathcal{N} =$ 4 SYM},'' {\em JHEP}, vol.~12, p.~122, 2016, arXiv:1601.05679.

\bibitem{BubbleAnsatz}
L.~F. Alday, D.~Gaiotto, and J.~Maldacena, ``{Thermodynamic Bubble Ansatz},''
  {\em JHEP}, vol.~09, p.~032, 2011, arXiv:0911.4708.

\bibitem{YsystemScatteringAmplitudes}
L.~F. Alday, J.~Maldacena, A.~Sever, and P.~Vieira, ``{Y-system for Scattering
  Amplitudes},'' {\em J. Phys.}, vol.~A43, p.~485401, 2010, arXiv:1002.2459.

\bibitem{OPEpolygonalWilsonLines}
L.~F. Alday, D.~Gaiotto, J.~Maldacena, A.~Sever, and P.~Vieira, ``{An Operator
  Product Expansion for Polygonal null Wilson Loops},'' {\em JHEP}, vol.~04,
  p.~088, 2011, arXiv:1006.2788.

\bibitem{SmatrixFiniteCoupling}
B.~Basso, A.~Sever, and P.~Vieira, ``{Spacetime and Flux Tube S-Matrices at
  Finite Coupling for N=4 Supersymmetric Yang-Mills Theory},'' {\em Phys. Rev.
  Lett.}, vol.~111, no.~9, p.~091602, 2013, arXiv:1303.1396.

\bibitem{OPEHelicityAmplitudes}
B.~Basso, J.~Caetano, L.~Cordova, A.~Sever, and P.~Vieira, ``{OPE for all
  Helicity Amplitudes},'' {\em JHEP}, vol.~08, p.~018, 2015, arXiv:1412.1132.

\bibitem{AsymptoticBetheAnsatzGKPvacuum}
D.~Fioravanti, S.~Piscaglia, and M.~Rossi, ``{Asymptotic Bethe Ansatz on the
  GKP vacuum as a defect spin chain: scattering, particles and minimal area
  Wilson loops},'' {\em Nucl. Phys.}, vol.~B898, pp.~301--400, 2015,
  arXiv:1503.08795.

\bibitem{adjointBFKL}
B.~Basso, S.~Caron-Huot, and A.~Sever, ``{Adjoint BFKL at finite coupling: a
  short-cut from the collinear limit},'' {\em JHEP}, vol.~01, p.~027, 2015,
  arXiv:1407.3766.

\bibitem{GromovBFKL1}
M.~Alfimov, N.~Gromov, and V.~Kazakov, ``{QCD Pomeron from AdS/CFT Quantum
  Spectral Curve},'' {\em JHEP}, vol.~07, p.~164, 2015, arXiv:1408.2530.

\bibitem{GromovBFKL2}
N.~Gromov, F.~Levkovich-Maslyuk, and G.~Sizov, ``{Pomeron Eigenvalue at Three
  Loops in $\mathcal N=$ 4 Supersymmetric Yang-Mills Theory},'' {\em Phys. Rev.
  Lett.}, vol.~115, no.~25, p.~251601, 2015, arXiv:1507.04010.

\bibitem{StructureConstantsPentagons}
B.~Basso, S.~Komatsu, and P.~Vieira, ``{Structure Constants and Integrable
  Bootstrap in Planar N=4 SYM Theory},'' 2015, arXiv:1505.06745.

\bibitem{StructureConstantsWrappingOrder}
B.~Basso, V.~Goncalves, and S.~Komatsu, ``{Structure constants at wrapping
  order},'' {\em JHEP}, vol.~05, p.~124, 2017, arXiv:1702.02154.

\bibitem{ClusteringThreePointFunctions}
Y.~Jiang, S.~Komatsu, I.~Kostov, and D.~Serban, ``{Clustering and the
  Three-Point Function},'' {\em J. Phys.}, vol.~A49, no.~45, p.~454003, 2016,
  arXiv:1604.03575.

\bibitem{StructureConstantsLightRayOperators}
I.~Balitsky, V.~Kazakov, and E.~Sobko, ``{Structure constant of twist-2
  light-ray operators in the Regge limit},'' {\em Phys. Rev.}, vol.~D93, no.~6,
  p.~061701, 2016, arXiv:1506.02038.

\bibitem{defectCFT1}
M.~de~Leeuw, C.~Kristjansen, and K.~Zarembo, ``{One-point Functions in Defect
  CFT and Integrability},'' {\em JHEP}, vol.~08, p.~098, 2015,
  arXiv:1506.06958.

\bibitem{defectCFT2}
I.~Buhl-Mortensen, M.~de~Leeuw, C.~Kristjansen, and K.~Zarembo, ``{One-point
  Functions in AdS/dCFT from Matrix Product States},'' {\em JHEP}, vol.~02,
  p.~052, 2016, arXiv:1512.02532.

\bibitem{defectCFT3}
I.~Buhl-Mortensen, M.~de~Leeuw, A.~C. Ipsen, C.~Kristjansen, and M.~Wilhelm,
  ``{One-loop one-point functions in gauge-gravity dualities with defects},''
  {\em Phys. Rev. Lett.}, vol.~117, no.~23, p.~231603, 2016, arXiv:1606.01886.

\bibitem{N4SYMQSC1}
N.~Gromov, V.~Kazakov, S.~Leurent, and D.~Volin, ``{Quantum Spectral Curve for
  Planar $\mathcal{N} =$ Super-Yang-Mills Theory},'' {\em Phys. Rev. Lett.},
  vol.~112, no.~1, p.~011602, 2014, arXiv:1305.1939.

\bibitem{N4SYMQSC2}
N.~Gromov, V.~Kazakov, S.~Leurent, and D.~Volin, ``{Quantum spectral curve for
  arbitrary state/operator in AdS$_{5}$/CFT$_{4}$},'' {\em JHEP}, vol.~09,
  p.~187, 2015, arXiv:1405.4857.

\bibitem{twistedN4SYMQSC}
V.~Kazakov, S.~Leurent, and D.~Volin, ``{T-system on T-hook: Grassmannian
  Solution and Twisted Quantum Spectral Curve},'' {\em JHEP}, vol.~12, p.~044,
  2016, arXiv:1510.02100.

\bibitem{N4SYMQSC3}
C.~Marboe and D.~Volin, ``{The full spectrum of AdS5/CFT4 I: Representation
  theory and one-loop Q-system},'' 2017, arXiv:1701.03704.

\bibitem{ABJMQSC}
A.~Cavaglià, D.~Fioravanti, N.~Gromov, and R.~Tateo, ``{Quantum Spectral Curve
  of the $\mathcal N=$ 6 Supersymmetric Chern-Simons Theory},'' {\em Phys. Rev.
  Lett.}, vol.~113, no.~2, p.~021601, 2014, arXiv:1403.1859.

\bibitem{ABJMQSCdetailed}
D.~Bombardelli, A.~Cavaglià, D.~Fioravanti, N.~Gromov, and R.~Tateo, ``{The
  full Quantum Spectral Curve for $AdS_4/CFT_3$},'' 2017, arXiv:1701.00473.

\bibitem{QSCetadeformed}
R.~Klabbers and S.~J. van Tongeren, ``{Quantum Spectral Curve for the
  eta-deformed AdS$_5$xS$^5$ superstring},'' {\em Nucl. Phys.}, vol.~B925,
  pp.~252--318, 2017, arXiv:1708.02894.

\bibitem{VolinPerturbativeSolution}
C.~Marboe and D.~Volin, ``{Quantum spectral curve as a tool for a perturbative
  quantum field theory},'' {\em Nucl. Phys.}, vol.~B899, pp.~810--847, 2015,
  arXiv:1411.4758.

\bibitem{ABJMQSC12loops}
L.~Anselmetti, D.~Bombardelli, A.~Cavaglià, and R.~Tateo, ``{12 loops and
  triple wrapping in ABJM theory from integrability},'' {\em JHEP}, vol.~10,
  p.~117, 2015, arXiv:1506.09089.

\bibitem{N4SYM1loop}
L.~N. Lipatov, ``{Next-to-leading corrections to the BFKL equation and the
  effective action for high energy processes in QCD},'' {\em Nucl. Phys. Proc.
  Suppl.}, vol.~99A, pp.~175--179, 2001.
\newblock [,175(2001)].

\bibitem{N4SYM2loop1}
A.~V. Kotikov and L.~N. Lipatov, ``{NLO corrections to the BFKL equation in QCD
  and in supersymmetric gauge theories},'' {\em Nucl. Phys.}, vol.~B582,
  pp.~19--43, 2000, hep-ph/0004008.

\bibitem{N4SYM2loop2}
A.~V. Kotikov, L.~N. Lipatov, and V.~N. Velizhanin, ``{Anomalous dimensions of
  Wilson operators in N=4 SYM theory},'' {\em Phys. Lett.}, vol.~B557,
  pp.~114--120, 2003, hep-ph/0301021.

\bibitem{N4SYM2loop3}
A.~V. Kotikov and L.~N. Lipatov, ``{DGLAP and BFKL evolution equations in the
  N=4 supersymmetric gauge theory},'' in {\em {35th Annual Winter School on
  Nuclear and Particle Physics Repino, Russia, February 19-25, 2001}}, 2001,
  hep-ph/0112346.

\bibitem{N4SYM2loop4}
A.~V. Kotikov and L.~N. Lipatov, ``{DGLAP and BFKL equations in the $N=4$
  supersymmetric gauge theory},'' {\em Nucl. Phys.}, vol.~B661, pp.~19--61,
  2003, hep-ph/0208220.
\newblock [Erratum: Nucl. Phys.B685,405(2004)].

\bibitem{N4SYM3loop}
A.~V. Kotikov, L.~N. Lipatov, A.~I. Onishchenko, and V.~N. Velizhanin, ``{Three
  loop universal anomalous dimension of the Wilson operators in $N=4$ SUSY
  Yang-Mills model},'' {\em Phys. Lett.}, vol.~B595, pp.~521--529, 2004,
  hep-th/0404092.
\newblock [Erratum: Phys. Lett.B632,754(2006)].

\bibitem{N4SYM4loop}
A.~V. Kotikov, L.~N. Lipatov, A.~Rej, M.~Staudacher, and V.~N. Velizhanin,
  ``{Dressing and wrapping},'' {\em J. Stat. Mech.}, vol.~0710, p.~P10003,
  2007, arXiv:0704.3586.

\bibitem{N4SYM5loop}
T.~Lukowski, A.~Rej, and V.~N. Velizhanin, ``{Five-Loop Anomalous Dimension of
  Twist-Two Operators},'' {\em Nucl. Phys.}, vol.~B831, pp.~105--132, 2010,
  arXiv:0912.1624.

\bibitem{N4SYM6looptwist3}
V.~N. Velizhanin, ``{Six-Loop Anomalous Dimension of Twist-Three Operators in
  N=4 SYM},'' {\em JHEP}, vol.~11, p.~129, 2010, arXiv:1003.4717.

\bibitem{N4SYM6loop}
C.~Marboe, V.~Velizhanin, and D.~Volin, ``{Six-loop anomalous dimension of
  twist-two operators in planar $ \mathcal{N}=4 $ SYM theory},'' {\em JHEP},
  vol.~07, p.~084, 2015, arXiv:1412.4762.

\bibitem{N4SYM7loop}
C.~Marboe and V.~Velizhanin, ``{Twist-2 at seven loops in planar $ \mathcal{N}
  $ = 4 SYM theory: full result and analytic properties},'' {\em JHEP},
  vol.~11, p.~013, 2016, arXiv:1607.06047.

\bibitem{Beccaria1}
M.~Beccaria and G.~Macorini, ``{QCD properties of twist operators in the N=6
  Chern-Simons theory},'' {\em JHEP}, vol.~06, p.~008, 2009, arXiv:0904.2463.

\bibitem{MinahanABJM4loop1}
J.~A. Minahan, O.~Ohlsson~Sax, and C.~Sieg, ``{Magnon dispersion to four loops
  in the ABJM and ABJ models},'' {\em J. Phys.}, vol.~A43, p.~275402, 2010,
  arXiv:0908.2463.

\bibitem{MinahanABJM4loop2}
J.~A. Minahan, O.~Ohlsson~Sax, and C.~Sieg, ``{Anomalous dimensions at four
  loops in N=6 superconformal Chern-Simons theories},'' {\em Nucl. Phys.},
  vol.~B846, pp.~542--606, 2011, arXiv:0912.3460.

\bibitem{ABJMspectroscopy}
G.~Papathanasiou and M.~Spradlin, ``{Two-Loop Spectroscopy of Short ABJM
  Operators},'' {\em JHEP}, vol.~02, p.~072, 2010, arXiv:0911.2220.

\bibitem{ABJMsuperspace}
M.~Leoni, A.~Mauri, J.~A. Minahan, O.~Ohlsson~Sax, A.~Santambrogio, C.~Sieg,
  and G.~Tartaglino-Mazzucchelli, ``{Superspace calculation of the four-loop
  spectrum in N=6 supersymmetric Chern-Simons theories},'' {\em JHEP}, vol.~12,
  p.~074, 2010, arXiv:1010.1756.

\bibitem{Beccaria2}
M.~Beccaria, F.~Levkovich-Maslyuk, and G.~Macorini, ``{On wrapping corrections
  to GKP-like operators},'' {\em JHEP}, vol.~03, p.~001, 2011, arXiv:1012.2054.

\bibitem{ABJM2}
M.~A. Bandres, A.~E. Lipstein, and J.~H. Schwarz, ``{Studies of the ABJM Theory
  in a Formulation with Manifest SU(4) R-Symmetry},'' {\em JHEP}, vol.~09,
  p.~027, 2008, arXiv:0807.0880.

\bibitem{KloseABJMreview}
T.~Klose, ``{Review of AdS/CFT Integrability, Chapter IV.3: N=6 Chern-Simons
  and Strings on AdS4xCP3},'' {\em Lett. Math. Phys.}, vol.~99, pp.~401--423,
  2012, arXiv:1012.3999.

\bibitem{StringDualN6ChernSimons}
G.~Grignani, T.~Harmark, and M.~Orselli, ``{The SU(2) x SU(2) sector in the
  string dual of N=6 superconformal Chern-Simons theory},'' {\em Nucl. Phys.},
  vol.~B810, pp.~115--134, 2009, arXiv:0806.4959.

\bibitem{hconjecture1}
N.~Gromov and G.~Sizov, ``{Exact Slope and Interpolating Functions in N=6
  Supersymmetric Chern-Simons Theory},'' {\em Phys. Rev. Lett.}, vol.~113,
  no.~12, p.~121601, 2014, arXiv:1403.1894.

\bibitem{hconjecture2}
A.~Cavaglià, N.~Gromov, and F.~Levkovich-Maslyuk, ``{On the Exact
  Interpolating Function in ABJ Theory},'' {\em JHEP}, vol.~12, p.~086, 2016,
  arXiv:1605.04888.

\bibitem{Ablinger1}
J.~Ablinger, {\em {A Computer Algebra Toolbox for Harmonic Sums Related to
  Particle Physics}}.
\newblock PhD thesis, Linz U., 2009, arXiv:1011.1176.

\bibitem{Ablinger2}
J.~Ablinger, {\em {Computer Algebra Algorithms for Special Functions in
  Particle Physics}}.
\newblock PhD thesis, Linz U., 2012-04, arXiv:1305.0687.

\bibitem{Ablinger3}
J.~Ablinger, J.~Blümlein, and C.~Schneider, ``{Analytic and Algorithmic
  Aspects of Generalized Harmonic Sums and Polylogarithms},'' {\em J. Math.
  Phys.}, vol.~54, p.~082301, 2013, arXiv:1302.0378.

\bibitem{Ablinger4}
J.~Ablinger, J.~Blumlein, and C.~Schneider, ``{Harmonic Sums and Polylogarithms
  Generated by Cyclotomic Polynomials},'' {\em J. Math. Phys.}, vol.~52,
  p.~102301, 2011, arXiv:1105.6063.

\bibitem{BlumleinStructuralRelations}
J.~Blumlein, ``{Structural Relations of Harmonic Sums and Mellin Transforms up
  to Weight w = 5},'' {\em Comput. Phys. Commun.}, vol.~180, pp.~2218--2249,
  2009, arXiv:0901.3106.

\bibitem{RemiddiVermaseren}
E.~Remiddi and J.~A.~M. Vermaseren, ``{Harmonic polylogarithms},'' {\em Int. J.
  Mod. Phys.}, vol.~A15, pp.~725--754, 2000, hep-ph/9905237.

\bibitem{Vermaseren}
J.~A.~M. Vermaseren, ``{Harmonic sums, Mellin transforms and integrals},'' {\em
  Int. J. Mod. Phys.}, vol.~A14, pp.~2037--2076, 1999, hep-ph/9806280.

\bibitem{LipatovSpinChain}
L.~N. Lipatov, ``{Asymptotic behavior of multicolor QCD at high energies in
  connection with exactly solvable spin models},'' {\em JETP Lett.}, vol.~59,
  pp.~596--599, 1994, hep-th/9311037.
\newblock [Pisma Zh. Eksp. Teor. Fiz.59,571(1994)].

\bibitem{FaddeevKorchemsky}
L.~D. Faddeev and G.~P. Korchemsky, ``{High-energy QCD as a completely
  integrable model},'' {\em Phys. Lett.}, vol.~B342, pp.~311--322, 1995,
  hep-th/9404173.

\bibitem{Kotikov1}
A.~V. Kotikov, A.~Rej, and S.~Zieme, ``{Analytic three-loop Solutions for N=4
  SYM Twist Operators},'' {\em Nucl. Phys.}, vol.~B813, pp.~460--483, 2009,
  arXiv:0810.0691.

\bibitem{Kotikov2}
M.~Beccaria, A.~V. Belitsky, A.~V. Kotikov, and S.~Zieme, ``{Analytic solution
  of the multiloop Baxter equation},'' {\em Nucl. Phys.}, vol.~B827,
  pp.~565--606, 2010, arXiv:0908.0520.

\bibitem{KotikovVeretin}
J.~Fleischer, A.~V. Kotikov, and O.~L. Veretin, ``{Analytic two loop results
  for selfenergy type and vertex type diagrams with one nonzero mass},'' {\em
  Nucl. Phys.}, vol.~B547, pp.~343--374, 1999, hep-ph/9808242.

\bibitem{Tolya1}
A.~V. Kotikov, ``{The Property of maximal transcendentality in the N=4
  Supersymmetric Yang-Mills},'' in {\em {In *Diakonov, D. (ed.): Subtleties in
  quantum field theory* 150-174}}, pp.~150--174, 2010, arXiv:1005.5029.

\bibitem{Tolya2}
A.~V. Kotikov, ``{The property of maximal transcendentality: calculation of
  Feynman integrals},'' {\em Theor. Math. Phys.}, vol.~190, no.~3,
  pp.~391--401, 2017, arXiv:1601.00486.

\bibitem{fishnet1}
A.~B. Zamolodchikov, ``{'Fishnet' diagrams as a completely integrable
  system},'' {\em Phys. Lett.}, vol.~97B, pp.~63--66, 1980.

\bibitem{Isaev}
D.~Chicherin, S.~Derkachov, and A.~P. Isaev, ``{Conformal group: R-matrix and
  star-triangle relation},'' {\em JHEP}, vol.~04, p.~020, 2013,
  arXiv:1206.4150.

\bibitem{fishnet2}
O.~Gürdoğan and V.~Kazakov, ``{New Integrable 4D Quantum Field Theories from
  Strongly Deformed Planar $\mathcal N = $ 4 Supersymmetric Yang-Mills
  Theory},'' {\em Phys. Rev. Lett.}, vol.~117, no.~20, p.~201602, 2016,
  arXiv:1512.06704.
\newblock [Addendum: Phys. Rev. Lett.117,no.25,259903(2016)].

\bibitem{fishnet3}
J.~Caetano, O.~Gurdogan, and V.~Kazakov, ``{Chiral limit of N = 4 SYM and ABJM
  and integrable Feynman graphs},'' 2016, arXiv:1612.05895.

\bibitem{fishnet4}
D.~Chicherin, V.~Kazakov, F.~Loebbert, D.~Müller, and D.-l. Zhong, ``{Yangian
  Symmetry for Bi-Scalar Loop Amplitudes},'' 2017, arXiv:1704.01967.

\bibitem{fishnet5}
B.~Basso and L.~J. Dixon, ``{Gluing Ladder Feynman Diagrams into Fishnets},''
  {\em Phys. Rev. Lett.}, vol.~119, no.~7, p.~071601, 2017, arXiv:1705.03545.

\bibitem{fishnet6}
N.~Gromov, V.~Kazakov, G.~Korchemsky, S.~Negro, and G.~Sizov, ``{Integrability
  of Conformal Fishnet Theory},'' 2017, arXiv:1706.04167.

\bibitem{fishnet7}
D.~Chicherin, V.~Kazakov, F.~Loebbert, D.~Müller, and D.-l. Zhong, ``{Yangian
  Symmetry for Fishnet Feynman Graphs},'' 2017, arXiv:1708.00007.

\end{thebibliography}

\end{document}